\documentclass[sigconf,twocolumn]{acmart}
\AtBeginDocument{%
  \providecommand\BibTeX{{%
    \normalfont B\kern-0.5em{\scshape i\kern-0.25em b}\kern-0.8em\TeX}}}

\usepackage{lipsum} 
\usepackage{xcolor}
\usepackage{adjustbox}
\usepackage{amsmath,amsfonts}
\usepackage{algorithmic}
\usepackage{graphicx}
\usepackage{caption}
\usepackage{subcaption}
\usepackage{textcomp}
\usepackage{todonotes}
\usepackage{xcolor}
\usepackage[table]{xcolor}
\usepackage{cleveref}
\usepackage[most]{tcolorbox}
\usepackage[inline]{enumitem}
\usepackage{multirow}
\usepackage{hyperref}
\usepackage{algorithmic}
\usepackage{stfloats}
\usepackage[font=normalsize]{caption}
\usepackage[font=bf]{caption}
\usepackage[utf8]{inputenc}

\tcbuselibrary{listingsutf8, skins, breakable}
\AtBeginDocument{%
  \providecommand\BibTeX{{%
  Bib\TeX}}}
\usepackage{listings}
\definecolor{codegreen}{rgb}{0,0.6,0}
\definecolor{codegray}{rgb}{0.5,0.5,0.5}
\definecolor{codepurple}{rgb}{0.58,0,0.82}
\definecolor{backcolour}{rgb}{0.95,0.95,0.92}
\lstdefinestyle{mystyle}{
    backgroundcolor=\color{backcolour},   
    commentstyle=\color{codegreen},
    keywordstyle=\color{magenta},
    numberstyle=\tiny\color{codegray},
    stringstyle=\color{codepurple},
    basicstyle=\ttfamily\footnotesize,
    breakatwhitespace=true,         
    breaklines=false,                 
    captionpos=b,                    
    keepspaces=true,                 
    numbers=left,                    
    numbersep=5pt,                  
    showspaces=false,                
    showstringspaces=false,
    showtabs=false,                  
    tabsize=2
}
\newcommand{\colorcircle}[2][1ex]{%
  \tikz[baseline=-.5ex] \fill[#2] (0,0) circle (#1);%
}

\newcounter{findingid}
\newcommand{\finding}[2]{%
  \refstepcounter{findingid}%
  \item[] % no automatic label
  \colorbox{blue!6}{%
    \parbox{\dimexpr\linewidth-2\fboxsep}{% ensures it spans nicely
      \textbf{Finding \arabic{findingid}:} \textit{#2}%
      \label{#1}%
    }%
  }%
}

\newcommand{\myred}[1]{\textbf{\textcolor{red}{#1}}}
\newcommand{\myblue}[1]{\textbf{\textcolor{blue}{#1}}}
\newcommand{\mygreen}[1]{\textbf{\textcolor{green!50!black}{#1}}}

\newcommand{\bc}[1]{%
  \tikz[baseline=(char.base)]{
    \node[shape=circle, fill=black, text=white, inner sep=1pt] (char) {#1};
  }%
}

\setcopyright{acmlicensed}
\copyrightyear{2026}
\acmYear{2026}
\setcopyright{cc}
\setcctype{by}
\acmConference[ICPE Companion '26]{Companion of the 17th ACM/SPEC International Conference on Performance Engineering}{May 04--08, 2026}{Florence, Italy}
\acmBooktitle{Companion of the 17th ACM/SPEC International Conference on Performance Engineering (ICPE Companion '26), May 04--08, 2026, Florence, Italy}
\acmDOI{10.1145/3777911.3801103}
\acmISBN{979-8-4007-2326-1/2026/05}

\begin{document}

% \copyrightyear{2026}
% \acmYear{2026}
% \setcopyright{cc}
% \setcctype{by}
% \acmConference[ICPE Companion '26]{Companion of the 17th ACM/SPEC International Conference on Performance Engineering}{May 04--08, 2026}{Florence, Italy}
% \acmBooktitle{Companion of the 17th ACM/SPEC International Conference on Performance Engineering (ICPE Companion '26), May 04--08, 2026, Florence, Italy}
% \acmDOI{10.1145/3777911.3801103}
% \acmISBN{979-8-4007-2326-1/2026/05}

%%
%% The "title" command has an optional parameter,
%% allowing the author to define a "short title" to be used in page headers.

% \title{Leveraging LLMs for Structured Information Extraction and Analysis from Cloud Incident Reports (Work In Progress Paper)}
\title[Leveraging LLMs for Structured Information Extraction and Analysis from Cloud Incident Reports \\ (Work In Progress Paper)]{Leveraging LLMs for Structured Information Extraction and Analysis from Cloud Incident Reports (Work In Progress Paper)}
% \title{Comparative Analysis of LLMs for Report Data Extraction}

%%
%% The "author" command and its associated commands are used to define
%% the authors and their affiliations.
%% Of note is the shared affiliation of the first two authors, and the
%% "authornote" and "authornotemark" commands
%% used to denote shared contribution to the research.

\author{Xiaoyu Chu}
\email{x.chu@vu.nl}
\affiliation{%
  \institution{Vrije Universiteit Amsterdam}
  \country{The Netherlands}
}

\author{Shashikant Ilager}
\email{s.s.ilager@uva.nl}
\affiliation{%
  \institution{Universiteit van Amsterdam}
  \country{The Netherlands}
}

\author{Yizhen Zang}
\email{y.zang-3@student.tudelft.nl}
\affiliation{%
  \institution{Delft University of Technology}
  \country{The Netherlands}
}

\author{Sacheendra Talluri}
\email{s.talluri@vu.nl}
\affiliation{%
  \institution{Vrije Universiteit Amsterdam}
  \country{The Netherlands}
}

\author{Alexandru Iosup}
\email{a.iosup@vu.nl}
\affiliation{%
  \institution{Vrije Universiteit Amsterdam}
  \country{The Netherlands}
}

% \affiliation{%
%   \institution{Institute for Clarity in Documentation}
%   \streetaddress{P.O. Box 1212}
%   \city{Dublin}
%   \state{Ohio}
%   \country{USA}
%   \postcode{43017-6221}
% }
%%
%% By default, the full list of authors will be used in the page
%% headers. Often, this list is too long, and will overlap
%% other information printed in the page headers. This command allows
%% the author to define a more concise list
%% of authors' names for this purpose.

% \renewcommand{\shortauthors}{Anonymous authors, et al.}

%%
%% The abstract is a short summary of the work to be presented in the
%% article.
\begin{abstract}
Incident management is essential to maintain the reliability and availability of cloud computing services. 
Cloud vendors typically disclose incident reports to the public, summarizing the failures and recovery process to help minimize their impact.
However, such reports are often lengthy and unstructured, making them difficult to understand, analyze, and use for long-term dependability improvements. %assessment. 
The emergence of LLMs offers new opportunities to address this challenge, but how to achieve this is currently understudied.
In this paper, we explore the use of cutting-edge LLMs to extract key information from unstructured cloud incident reports. 
First, we collect more than 3,000 incident reports from 3 leading cloud service providers (AWS, AZURE, and GCP), and manually annotate these collected samples.
Then, we design and compare 6 prompt strategies to extract and classify different types of information. 
We consider 6~LLM models, including 3 lightweight and 3 state-of-the-art (SotA), and evaluate model accuracy, latency, and token cost across datasets, models, prompts, and extracted fields.
% Finally, we identify the best-performing prompt strategy and model to extract information from all reports and perform a statistical dependability analysis.
%%% CHECK FROM HERE ON, AFTER ALL THE RESULTS ARE IN
%Our results show that the best average accuracies for the AWS, AZURE, and GCP datasets are 0.90, 0.79, and 0.74, respectively.
Our study has uncovered the following key findings:
% (1) The average accuracy of LLM extraction of metadata fields is high, over 75\% and up to 95\% depending on the dataset.
(1) LLMs achieve high metadata extraction accuracy, $75\%\text{--}95\%$ depending on the dataset.
(2) Few-shot prompting generally improves accuracy for meta-data fields except for classification, and has better (lower) latency due to shorter output-tokens but requires $1.5\text{--}2\times$ more input-tokens.
(3) Lightweight models (e.g., Gemini~2.0, GPT~3.5) offer favorable trade-offs in accuracy, cost, and latency; SotA models yield higher accuracy at significantly greater cost and latency.
% (4) \red{TODO: MTTR or other result of the dependability analysis in §5}.
% (4) Using the best-performing model on the full AZURE dataset, we find that DEPLOY, CONFIG, and EXTERNAL are the 3 most common root causes, followed by OVERLOAD and MAINTAIN.
 % (4) Incident durations vary by cloud vendor; Azure incidents last $2.6\times$ longer than AWS.
Our study provides tools, methodologies, and insights for leveraging LLMs to accurately and efficiently extract incident-report information.
The FAIR data and code are publicly available at \url{https://github.com/atlarge-research/llm-cloud-incident-extraction}.
\end{abstract}

% \begin{CCSXML}
% <ccs2012>
% <concept>
% <concept_id>10010520.10010575.10010577</concept_id>
% <concept_desc>Computer systems organization~Reliability</concept_desc>
% <concept_significance>500</concept_significance>
% </concept>
% </ccs2012>
% \end{CCSXML}

% \ccsdesc[500]{Computer systems organization~Reliability}

\maketitle

\section{Introduction}\label{sec:intro}

% 1) Context and problem
% 2) Related work and why can’t it solve the problem
% 3) Use cases of findings with references
% 4) 1 Major research questions (high-level) + 3 small research
% questions (Why important? Why novel? Why
% challenging? Why useful?)
% 5) Approaches for each question
% 6) Expected contributions for each question

% \textbf{\textit{Context:}}
Incident management is an important and complex process in cloud computing operations~\cite{7027595, 10.1145/3663529.3663841, DBLP:conf/icse/AhmedGBZZR23}. 
A cloud incident could occur for multiple reasons, including hardware or software failures, deployment and runtime errors, network disruptions, or security breaches~\cite{DBLP:conf/cloud/GunawiHSLSAE16, 10.1145/3510457.3513030}.
% \todo[inline]{Shashi: Define what is incident here, then incident report.
% For instance: An incident in the cloud could occur for multiple reasons, including hardware or software failures, deployment and runtime errors, network disruptions, or security breaches, among others ~\cite{}. } 
Such incidents can disrupt user services, affecting the reliability of the system. To minimize the impact, once an incident occurs, cloud operators provide public incident reports~\cite{report-aws, report-azure, report-gcp} to inform users about the incident management process and root causes.
% \footnote{}.
However, these reports are very complex, lengthy, and too unstructured to understand or use for long-term analysis and mitigation. 
As a result, while individual reports are disclosed for each incident, there is a lack of a public dataset and long-term analysis of these incidents.
Currently, LLMs have become powerful tools and offer new opportunities to address such problem. 
% \textcolor{red}{TODO: add llm applications for system engineering and performance engineering}
LLMs have already been effectively used in performance engineering, such as log parsing~\cite{10.1145/3674805.3686684, 10.1145/3643916.3644408}, root cause analysis~\cite{DBLP:conf/icse/AhmedGBZZR23, DBLP:conf/cikm/WangLZZWYF0W24, 10.1145/3663529.3663841, DBLP:conf/eurosys/ChenXMKGSCGFWZG24, DBLP:conf/sigsoft/GoelHSGPBZR24}. 
% Among those LLM applications, information extraction is a key task in natural language processing (NLP), where structured information is automatically extracted from unstructured text. 
Therefore, we propose to leverage advanced LLMs to extract and analyze key information from public cloud incident reports.

% \begin{figure}[t]
%   \centering
%   \includegraphics[width=\linewidth]{figure/figure1-example-extraction.drawio.pdf}
%   \vspace{-0.8cm}
%   % \caption{Illustration of incident report data extraction, from public cloud operators to structured data.}
%   %\caption{Structure of this work.}
%   \caption{LLM-based extraction of structured data from incident reports, and incident analysis. This work collects incident reports~($\S$2), publishes structured data (label ``D''), and contributes to building the workflow ($\S$3-5).}
%   \vspace{-0.5cm}
%   \label{fig:1}
% \end{figure}

% \textbf{\textit{Challenges:}}
% \todo[size=\tiny]{Keep \Cref{fig:1} and the start of its explanation on page 1.}
We design a workflow to adapt LLMs for report data extraction.
% address three main challenges, 
% \Cref{fig:1} illustrates the use of LLMs to extract structured data from incident reports, and then to conduct semi-automated incident analysis on the structured data. 
Our work addresses three main challenges in this workflow process:
First, \textbf{there is no publicly available labeled dataset of cloud incident reports, nor open source tools for automatically extracting and analyzing their key information.}
% , making it difficult to perform statistical characterization on these reports.
Previous studies on cloud incident and outage analysis have relied on manual or rule-based data extraction~\cite{DBLP:conf/cloud/GunawiHSLSAE16, DBLP:journals/corr/abs-2504-09476}, which is highly time-consuming and labor-intensive due to the length and complexity of the reports. 
For example, in our collected datasets from Azure and GCP, the average report length exceeds 500 words (see \Cref{tab:data}).
Second, \textbf{there is a lack of methodology and systematic evaluation for assessing the performance and cost of applying LLMs to incident report data extraction. } Although LLMs are widely used for information extraction~\cite{DBLP:conf/icse/ShettyBKRN021, DBLP:journals/corr/abs-2312-17617}, their design and application for incident report extraction remain unclear. In particular, it is not well understood how to structure data, how to design prompts, and how to adapt LLMs for cloud incident analysis.
% In particular, it is not well understood which models achieve the best accuracy, or how they compare in terms of latency and cost.
% although some studies have attempted to use advanced LLMs for information extraction
Third, \textbf{there are limited studies to perform long-term statistical characterization on incident reports.} 
Previous studies have conducted incident report analysis on cloud services~\cite{DBLP:conf/cloud/GunawiHSLSAE16, DBLP:journals/corr/abs-2504-09476},  
% or LLM services~\cite{10.1145/3676151.3719372, 10.1145/3680256.3721320}, 
but these analyses are typically based on labeled datasets and cover only limited fields, making them difficult to generalize to long-term datasets or diverse incident reports. 

Addressing these challenges, our key contributions are:
\begin{enumerate}
% [label=\textbf{\Roman*.}]
    \item \textbf{Dataset and toolbox construction (\Cref{sec:data}):} 
    This work is the first to provide open-sourced datasets and toolbox for systematically exploring and evaluating LLMs for structured information extraction and analysis of cloud incident reports.
    We collect over 3,000 cloud incident reports with annotated subsets for report-data evaluation. We also develop a toolbox for leveraging LLMs for report data extraction, evaluation, and analysis. 
    % We collect over 3,000 cloud incident reports from 3 leading cloud vendors, AWS, AZURE, GCP, and create annotated subsets for report-data extraction. We also develop a toolbox for leveraging LLM for report data extraction, evaluation, and analysis. 
    
    \item \textbf{LLMs adaption and evaluation (\Cref{sec:experiment} and \ref{sec:result}):} 
    This work demonstrates the usefulness of SotA LLMs such as GPT-4o for incident information extraction and analysis. 
    We systematically evaluate 6 LLMs models of different size with diverse prompt strategies, we compare their accuracy and cost with different evaluation metrics. Our results achieve 75\%–95\% accuracy on metadata extraction, based on the dataset. We also provide recommendations for selecting prompts and LLMs based on their latency and cost.
    % We compare 6 models. We design zero-shot-CoT and few-shot-CoT prompts. We systematically compare the performance of different LLM approaches with lexical and semantic accuracy metrics. We also compare their cost through token usage and latency. 
    
    % \item \textbf{Incident analysis (\Cref{sec:analysis}):} We conduct failure analysis based on the LLM-extracted data from cloud incident reports, including the duration of incidents, the categories of services affected by incidents, and root cause categories. We find that DEPLOY, CONFIG, and EXTERNAL are the three most popular root causes of cloud incidents.
    
    \item \textbf{Open science:} We follow the FAIR principles of open science and release the datasets\footnote{Zenodo: \url{https://zenodo.org/records/14010282}} and software\footnote{GitHub: \url{https://github.com/atlarge-research/llm-cloud-incident-extraction}} to enable reproducibility and further research.
\end{enumerate}

% \input{sections/2_background}
% \newpage
\vspace*{-0.25cm}
\section{Data Collection and Preparation}\label{sec:method}\label{sec:data}

% To enable an LLM to comprehend the problem context and produce meaningful analysis, it is essential to curate a diverse and well-structured dataset. 
% In our case, the dataset should enable prompting techniques, such as zero-shot and few-shot prompting, for LLMs to learn domain-specific knowledge. 
% Evaluating LLM performance in incident data extraction requires a labeled and well-structured dataset.
% In this section, we present the methodology used to construct such a dataset, beginning with data collection, followed by processing and annotation procedures.

% \todo[inline,size=\tiny]{What is novel? The collection process? The processing?// The raw data? The annotated? Etc.}
% \todo[inline,size=\tiny]{What is outstanding or surprising, overall?}

\subsection{Dataset Collection}\label{sec:method:dataset}\label{sec:dataset}
\begin{table}[t]
\centering
\caption{Summary of cloud incident reports.}
\vspace{-0.3cm}
\label{tab:data}\label{tab:dataset}
\resizebox{\linewidth}{!}{
\begin{tabular}{cccrrr}
\toprule
\textbf{ID} & \textbf{Name} & \textbf{Period} & \multicolumn{1}{c}{\textbf{\# Rows}} & \multicolumn{1}{l}{\textbf{\# Labeled}} & \multicolumn{1}{c}{\textbf{Avg. Words}} \\ \midrule
1           & AWS           & 2016 - 2022     & 774                                  & 150 (19\%)                                    & 151                                     \\
2           & AZURE         & 2019 - 2024     & 127                                  & 95 (75\%)                                     & 575                                     \\
3           & GCP           & 2016 - 2021     & 2,186                                & 215 (10\%)                                    & 533                                     \\ 
\midrule
TOTAL           & 3 sources           & 2016 - 2024     & 3,087                                & 460 (15\%)                                     & ---                                     \\ 
\bottomrule
\end{tabular}}
\vspace{-0.3cm}
\end{table}

\begin{table}[t]
\centering
\caption{Summary of extracted fields. 
% including type, availability and access method, and evaluation metric. 
Legend: \colorcircle[0.8ex]{blue} indicates fully extracted from the report, \colorcircle[0.8ex]{red} indicates inferred by LLMs. 
EM $=$ Exact Match, TK $=$ Token-Level F1, BS $=$ BERTScore.}
\vspace{-0.3cm}
\label{tab:extract-fields}\label{tab:evaluate-metric}
\resizebox{\linewidth}{!}{
\begin{tabular}{cllcccccl}
\toprule
\multirow{2}{*}{\textbf{ID}} & \multicolumn{1}{c}{\multirow{2}{*}{\textbf{Extracted Fields}}} & \multicolumn{1}{l}{\multirow{2}{*}{\textbf{Type}}} & \multicolumn{3}{c}{\textbf{Datasets}}                                                                                                                     & \multicolumn{3}{c}{\textbf{Eval. Metric}}                                                  \\ \cline{4-9} 
                    & \multicolumn{1}{c}{}                                  & \multicolumn{1}{c}{}                      & \textbf{AWS}                                            & \textbf{AZURE}                                          & \textbf{GCP}                                            & \textbf{EM}                        & \textbf{TK}                        & \multicolumn{1}{c}{\textbf{BS}}    \\ \midrule
1                   & Service Name                                          & entity                                    & \colorcircle[0.8ex]{blue} & \colorcircle[0.8ex]{blue} & \colorcircle[0.8ex]{blue}                                              & \checkmark &                           &                           \\
2                   & Location                                              & entity                                    & \colorcircle[0.8ex]{blue} & \colorcircle[0.8ex]{blue} & -                                              & \checkmark &                           &                           \\
3                   & Start Time                                            & entity                                    & \colorcircle[0.8ex]{blue} & \colorcircle[0.8ex]{blue} & \colorcircle[0.8ex]{blue} & \checkmark &                           &                           \\
4                   & End Time                                              & entity                                    & \colorcircle[0.8ex]{blue} & \colorcircle[0.8ex]{blue} & \colorcircle[0.8ex]{blue} & \checkmark &                           &                           \\
5                   & Timezone                                              & entity                                    & \colorcircle[0.8ex]{blue} & \colorcircle[0.8ex]{blue} & \colorcircle[0.8ex]{blue} & \checkmark &                           &                           \\
6                   & Service Categ.                                      & class                                     & \colorcircle[0.8ex]{red}  & \colorcircle[0.8ex]{red}  & \colorcircle[0.8ex]{red}  & \checkmark &                           &                           \\
7                   & Root Cause Categ.                                   & class                                     & -                                              & \colorcircle[0.8ex]{red}  & -                                              & \checkmark &                           &                           \\
8                   & User Symp. Categ.                                   & multi-class                               & \colorcircle[0.8ex]{red}  & \colorcircle[0.8ex]{red}  & \colorcircle[0.8ex]{red}  &                           & \checkmark &                           \\
9                   & User Symptom                                          & text                                      & \colorcircle[0.8ex]{blue} & \colorcircle[0.8ex]{blue} & \colorcircle[0.8ex]{blue} &                           &  & \checkmark \\
10                  & Root Cause                                            & text                                      & -                                              & \colorcircle[0.8ex]{blue} & -                                              &                           &  & \checkmark \\ \bottomrule
\end{tabular}}
\vspace{-0.5cm}
\end{table}

% \todo[inline,size=\tiny]{What is outstanding or surprising about data collection? The scale of the dataset (incidents)? The diversity of sources? Etc. // AI: I have already labeled the three sources 'leading [cloud providers]', which emphasizes the relevance of the data.//SI: Added textual description about diversity and structure}

Incident reports from cloud service providers are frequently published online via their respective web platforms~\cite{report-aws, report-azure, report-gcp}. However, these reports vary in structure, and the extent of publicly available information differs across providers. To compile a diverse set of incident report data, we sourced reports from 3 major cloud service providers: AWS, AZURE, and GCP, covering the period from 2016 to 2024. In total, we collected approximately 3,000 incident reports. Each report consists of important information such as the name of the service, the location, the user symptom, and the root cause.
% \todo[inline, size=tiny]{SI: 1-2 lines about data, may be some field from table 2 can be described here}

To \textit{collect} the data, we scraped and archived the incident report web pages from the 3 cloud providers. We then \textit{processed} the data using the methodology described in~\Cref{sec:data:annot}. Additionally, we \textit{manually labeled} approximately 15\% of the reports to serve as ground truth for evaluation. Although the proportion of labeled data varies by provider, the resulting sample size yields statistically meaningful insights; further details are provided in~\Cref{sec:annot}. 
%A summary of the cloud incident report dataset used in this study is presented in~\Cref{tab:data}.
\Cref{tab:data} presents a summary of this study's datasets.

%Due to the unbalanced numbers of reports, we \textit{manually labeled} about 15\% of approximately 20\%, 75\%, and 10\% of each cloud's reports as ground truth for evaluation. 
%%The detailed annotation process is described in~\Cref{sec:annot}.

\vspace*{-0.25cm}
\subsection{Data Processing} \label{sec:data:processing}

% In this stage, we process the raw dataset to prepare it for prompt-based techniques. 
In this stage, we process the raw dataset to prepare it for prompt construction. 
% A key challenge in working with incident reports is the inconsistency in the availability of information fields~(see \Cref{tab:extract-fields}) across different cloud service providers. For example, AWS reports typically omit root cause information, which introduces an additional classification task specific to AWS prompts. This absence of standardized fields complicates the extraction process and necessitates the design of customized prompt templates tailored to each provider. Consequently, we design customized prompt templates for each provider to ensure meaningful information can be extracted from operator incident reports to support further analysis. 
\Cref{tab:extract-fields} presents the target fields identified for extraction. These fields are categorized into entities, classes, and free-text segments, as each type requires distinct evaluation strategies, which are discussed in detail in~\Cref{sec:eval-metrics}. Accordingly, our data processing includes the following steps: 
(1) \textit{Raw datasets}: We extract information from each externally scraped file in HTML format and save it as a \emph{Parquet} file for each operator. (2) \textit{Cleaned datasets}: Because the structure of each vendor's report is different, we processed them separately to filter out invalid metrics (e.g., \verb|None| values and \verb|duplicates|). After this, we obtained cleaned datasets with an identical structure, which can be easily combined for unified analysis. (3) \textit{Sample datasets}: Because of scale, labeling the entire dataset is impractical, so we applied \emph{K-means} clustering to select representative sample datasets for data annotation. (4) \textit{Labeled datasets}: We manually annotated the fields listed in~\Cref{tab:extract-fields} from the sample datasets. Following a strict data annotation process (see~\Cref{sec:annot}), we obtained the labeled datasets.

\begin{figure}[t]
  \centering
  \includegraphics[width=\linewidth]{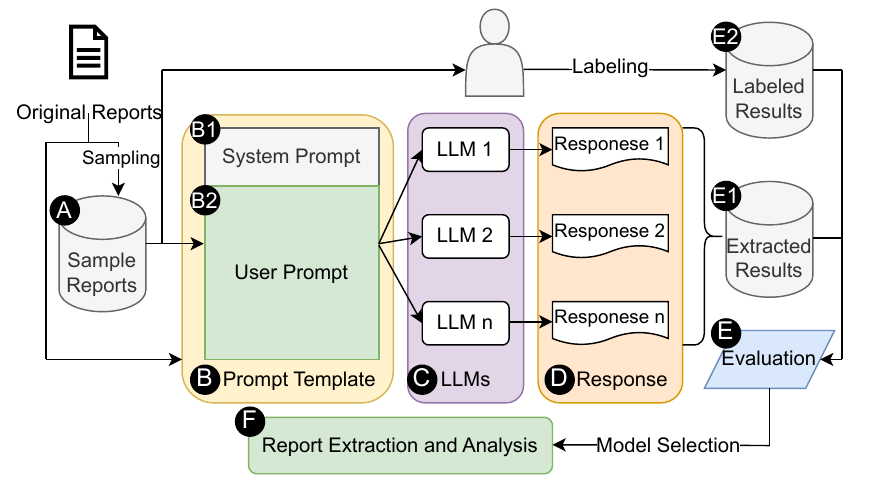}
  \vspace{-0.8cm}
  \caption{Overview of the workflow (structured method) for LLM-based data extraction from cloud incident reports.}
  \vspace{-0.5cm}
  \label{fig:overview}
\end{figure}

\vspace*{-0.25cm}
\subsection{Data Annotation}\label{sec:annot}\label{sec:data:annot}

% \todo[inline, size=tiny]{SI: define what annotation is in this context, and the importance of it for which process, then the need for manual annotations}

% \begin{table}[t]
% \centering
% \caption{Sub-classifications of services, user symptoms, and root causes. 
% The table lists the sub-classification tags that we manually defined and labeled from incident reports. 
% }
% \vspace{-0.3cm}
% \label{tab:sub-class}
% \resizebox{\linewidth}{!}{
% \begin{tabular}{ll}
% \toprule
% \textbf{Class} & \textbf{Sub-classification Tags}                                                                                                            \\ \toprule
% Service  Categories       & \begin{tabular}[c]{@{}l@{}}COMPUTE, STORAGE, NETWORK, SECURITY, \\ AI, MANAGEMENT, ANALYTICS, DATABASE, \\ OTHERS, and UNKNOWN\end{tabular} \\ \midrule
% User Symptoms     & ERROR, UNAVIL, DELAY, DEPERF, and OTHERS                                                                                                        \\ \midrule
% Root Causes     & \begin{tabular}[c]{@{}l@{}}CONFIG, OVERLOAD, DEPLOY, EXTERNAL, \\ MAINTAIN, OTHERS, and UNKNOWN\end{tabular}                                             \\ \bottomrule
% \end{tabular}}
% \vspace{-0.3cm}
% \end{table}

Data annotation is a crucial step, as ground truth data is necessary to compare against the LLMs' extracted results and to assess their accuracy.
Since no annotated cloud incident datasets are publicly available for evaluation, we created our own annotated datasets.
To ensure annotation accuracy and consistency, we selected 3 researchers with a background in computer systems as annotators.
% \begin{itemize}
%     \item \textbf{Annotator 1}: A doctoral researcher in distributed systems specializing in system modeling and analysis, with extensive experience in incident failure recovery processes.
%     \item \textbf{Annotator 2}: A graduate student in computer science specializing in software engineering, with experience in workload analysis and prediction.
%     \item \textbf{Annotator 3}: A computer system researcher specializing in datacenter reliability and sustainability, responsible for ensuring the annotation criteria. 
% \end{itemize}
The annotation process is as follows: 
First, Annotators 1 and 2 were asked to independently label the sample datasets following the provided instructions. 
After completing the labeling, they compared their results to check for alignment. In cases of disagreement, they had discussions to reach a consensus. 
If no agreement was possible, Annotator 3 intervened to make the final decision.

\section{Setup of LLM Experiments}\label{sec:experiment}\label{sec:exp:setup}

\begin{table}[t]
\centering
\caption{Model configurations. 
%Prices in \$ per 1M tokens. 
Model Type: L--lightweight model; S--state of the art. Models marked with ($^{*}$) do not provide time fingerprints, but the dates when these models were run are shown in \Cref{sec:result}. Price: T=Token.}
\vspace{-0.3cm}
\label{tab:model-config}
\resizebox{\linewidth}{!}{
\begin{tabular}{llrr}
\toprule
\multirow{2}{*}{\textbf{Model Alias}} & \multirow{2}{*}{\textbf{Model Type / Name}} & \multicolumn{2}{c}{\textbf{Price [\$/$10^6$T]}} \\
                                      &                                             & \textbf{Input}     & \textbf{Output}     \\ \midrule
GPT 4o                                & S / gpt-4o$^{*}$                                                                                  & 2.50                               & 10.00                               \\
GPT 3.5                               & L / gpt-3.5-turbo$^{*}$                                                                           & 0.50                               & 1.50                                \\
\midrule
Claude Sonnet 4                       & S / claude-sonnet-4-20250514                                                                & 3.00                               & 15.00                               \\
Claude 3.5                            & L / claude-3-5-haiku-20241022                                                               & 0.80                               & 4.00                                \\
\midrule
Gemini 2.5                            & S / gemini-2.5-pro$^{*}$                                                                          & 1.25                               & 10.00                               \\ 
Gemini 2.0                            & L / gemini-2.0-flash$^{*}$                                                                        & 0.10                               & 0.40                                \\
\bottomrule
\end{tabular}}
\vspace{-0.3cm}
\end{table}

% \begin{table}[t]
% \centering
% \caption{LLM API configurations. Prices per 1M tokens in \$.}
% \vspace{-0.3cm}
% \label{tab:llm-config}
% \resizebox{0.8\linewidth}{!}{
% \begin{tabular}{llll}
% \toprule
% \multirow{2}{*}{\textbf{Model}} & \multirow{2}{*}{\textbf{Temperature}} & \multicolumn{2}{c}{\textbf{Token Price}}                              \\
%                        &                              & \multicolumn{1}{c}{\textbf{Input}} & \multicolumn{1}{c}{\textbf{Output}} \\ \midrule
% GPT-3.5-turbo          & 0                            & 0.5                       & 1.5                        \\
% Claude-3-5-haiku       & 0                            & 0.25                      & 1.25                       \\
% Gemini-1.5-flash       & 0                            & 0.0375                    & 0.15                       \\ \bottomrule
% \end{tabular}}
% \end{table}

\vspace*{-0.2cm}
\begin{table}[t]
\centering
\caption{Six prompt strategies with different components. Acronyms: ZS=Zero-Shot; FS=Few-shot.}
\vspace{-0.3cm}
\label{tab:prompt-strategy}
\resizebox{\linewidth}{!}{
\begin{tabular}{ll}
\hline
\textbf{Label} & \textbf{Components}                       \\ \hline
Full-ZS        & Task + CoT + Category + Format            \\
Full-FS        & Task + CoT + Category + Examples + Format \\
Basic-ZS       & Task + Format                             \\
Basic-FS       & Task + Examples + Format                  \\
CoT-ZS         & Task + CoT + Format                       \\
Categ-ZS        & Task + Category + Format                  \\ \hline
\end{tabular}}
\vspace*{-0.5cm}
\end{table}

\begin{figure}[t]
\centering
% System Prompt
% \begin{tcolorbox}[title=System Prompt, colback=gray!5!white, colframe=gray!60!black, sharp corners]
% You are a system operator help to extract useful information from an incident report. 
% \end{tcolorbox}
% \vspace{-0.5cm}

% User Prompt
\begin{tcolorbox}[title=Task Description (Task), colback=black!5!white, colframe=black!60!black,  sharp corners]
Analyze the incident report step by step to extract structured information. 
\end{tcolorbox}
\vspace{-0.5cm}

\begin{tcolorbox}[title=Chain-of-Thought Instruction (CoT), colback=black!5!white, colframe=black!60!black,  sharp corners]

Follow the reasoning steps:

1. Identify the service name and service location.

2. From \verb|{service_category_lst}|, select one most relevant service category.

3. Extract the relevant sentence(s) that describe user symptoms. Then, from \verb|{user_symp_lst}|, select one or more categories that best match the extracted symptoms.
% with instruction: {\verb|{user_symp_instruction}|}.

4. Identify the start time, end time, and timezone. Format times as "HH:MM:SS" (24-hour).

\end{tcolorbox}
\vspace{-0.5cm}
\begin{tcolorbox}[title=Categorization Instruction (Category), colback=black!5!white, colframe=black!60!black, sharp corners]

The definition for user symptom category are: 

{\verb|{user_symp_instruction}|}.
\end{tcolorbox}
\vspace{-0.5cm}

\begin{tcolorbox}[title=Answering Format (Format), colback=black!5!white, colframe=black!60!black, sharp corners]
Finally, return the extracted information in a JSON object with the keys:

% \begin{verbatim}
$service\_name, location, service\_category, start\_time,$ 
$end\_time, timezone, user\_symptom, user\_symptom\_category.$
% \end{verbatim}

% Now, analyze the following incident report: 

% \textit{(Original report...)}

% title: \verb|{input_title}|,
% status: \verb|{input_status}|,
% description(HTML): \verb|{input_desc}|

\end{tcolorbox}
\vspace{-0.5cm}
% Few-shot Examples
\begin{tcolorbox}[title=Few-shot Examples (Examples), colback=black!5!white, colframe=black!60!black, sharp corners]

Here are a few examples of report content (labeled \textbf{Q}) and extracted information (labeled \textbf{A}).

\textbf{Q:} title: \myblue{Amazon CloudWatch} (\myblue{Ireland})

status: [RESOLVED] Delayed CloudWatch Metrics

description: 
% 11:55 AM PST We can confirm increased delays for CloudWatch log event processing for metric filter extraction and log subscriptions in the EU-WEST-1 Region. CloudWatch alarms may transition into "INSUFFICIENT\_DATA" state if set on metrics extracted using log filters. We are are working towards resolution.
% 12:49 PM PST We can confirm increased delays for CloudWatch log event processing for metric filter extraction and log subscriptions in the EU-WEST-1 Region. CloudWatch alarms may transition into "INSUFFICIENT\_DATA" state if set on metrics extracted using log filters. We have isolated the likely root cause to a subsystem that saw an unexpected jump in resource consumption. We continue to work towards resolution.
% 1:52 PM PST We have implemented a fix to address the CloudWatch log event processing delays in the EU-WEST-1 Region and are starting to see signs of recovery. We will provide an update once full recovery has been observed.
(...)
2:56 PM PST Between \myblue{10:26 AM} and \myblue{02:40 PM} PST, \myblue{we experienced increased delays for CloudWatch log event processing for metric filter extraction and log subscriptions in the EU-WEST-1 Region}. 
(...)
% This was due to a subsystem that saw an unexpected jump in resource consumption. The issue has been resolved and the service is operating normally. New events are processing as normal, while we work through the message backlog. We expect to completely drain the backlog over the next 1 hour.

\textbf{A:}
% \begin{verbatim}
    {"service\_name": "\myblue{Amazon CloudWatch}", 
     "location": "\myblue{Ireland}", 
     "service\_category": "\myred{management}", 
     "start\_time": "\myblue{10:26:00}", 
     "end\_time": "\myblue{14:40:00}", 
     "timezone": "\myblue{PST}", 
     "user\_symptom\_category": "\myred{DELAY}", 
     "user\_symptom": "\myblue{we experienced increased delays for CloudWatch log event processing for metric filter extraction and log subscriptions in the EU-WEST-1 Region.}"} 
% \end{verbatim}
%
\textit{(Continued...)}
\end{tcolorbox}

%\vspace{-0.35cm}
\vspace*{-0.5cm}
\caption{Full-FS user prompt template for extracting data from AWS reports, with five components: \emph{Task}, \emph{CoT}, \emph{Category}, \emph{Format}, and \emph{Examples}. Colors: \myblue{blue}=extracted, \myred{red}=classified.}
%%The input prompt is combined \emph{System Prompt}, \emph{Few-shot Examples}, and \emph{User Prompt}.}
% \myblue{myblue} text means directly extract while \myred{red} means classifications.}
\vspace{-0.7cm}
\label{fig:prompt-template}
\end{figure}

% We design and conduct LLM experiments as illustrated in \Cref{fig:overview}. The workflow emphasizes three key aspects: selecting appropriate models, developing effective prompt designs, and evaluating performance and cost metrics to assess model efficiency.
%model selection, prompt design (how the LLMs are used), and evaluation metrics for assessing model performance and cost.

%\subsection{Overview of LLMs for Report Extraction} \label{sec:exp:setup:llm}
\subsection{Use of LLMs for Report Extraction} \label{sec:exp:setup:llm}

\Cref{fig:overview} provides a methodological overview of this study.
%the methodology used in this study. 
Following data collection and annotation, the process moves to using LLMs for report extraction on sample datasets~(\bc{A}). 
First, we design a prompt template~(\bc{B}) that includes general context about cloud incidents and step-by-step instructions for data extraction, covering both zero-shot and few-shot settings. 
\Cref{fig:prompt-template} depicts an example full template, which contains five components.
The template varies slightly depending on the cloud operator, as their reports differ in structure and available information. 

For each report, we insert its content into the template as part of the user prompt~(\bc{B2}).
% which the LLMs run. 
A system prompt~(\bc{B1}) is set for LLMs to assume the role of a system operator to help perform data extraction from cloud incident reports.
Combined with system prompt, the complete prompt is then sent to a set of candidate LLMs~(\bc{C}), and the model responses~(\bc{D}) are collected in JSON format.
We evaluate~(\bc{E}) the accuracy of the extracted results~(\bc{E1}) by comparing them with the labeled datasets~(\bc{E2}), which serve as the ground truth. 
We also analyze and assess the latency and cost of candidate LLMs.
Finally, we select the best-performing LLMs for each operator and conduct incident report extraction and analysis~(\bc{F}).

% \todo[inline,size=\tiny]{ADD circular labels to \Cref{fig:overview} and use them in the text.}

\subsection{Environmental Setup and LLM Models} \label{sec:exp:setup:env}
\vspace{0.4cm}

% \input{table/table-env}

% \Cref{tab:env} outlines the experimental environment. 
% All experiments were conducted on a high-performance computing server equipped with 2 Intel Xeon Silver processors, each providing 10 cores per CPU. No GPU was used in the experiments.

\Cref{tab:model-config} presents the models selected in our experiments for comparison.
We selected models from popular LLM providers (OpenAI, Claude, and Gemini). 
% For each LLM API, we included both lightweight models (GPT-3.5, Claude~3.5, and Gemini~2.0) and top-tier models with state-of-the-art performance (GPT-4o, Claude~Sonnet~4, and Gemini~2.5~Pro) based on the updated LLM benchmark~\cite{lm-arena}. 
% The temperature parameter was set to 0 for all experiments.
To estimate cost per model, \Cref{tab:model-config} also lists the official-website pricing of input and output tokens. %, as provided on the official websites. 

% \todo[inline,size=\tiny]{Unclear what the temperature parameter is and why it is relevant---we are not referring to it later in the study, and we do not vary its value. Consider to omit this parameter, also from \Cref{tab:model-config}.}

% \input{table/table-env}

% \input{table/table-model-selections}

\vspace*{-0.25cm}
\subsection{Prompt Engineering} \label{sec:exp:setup:prompt}

% \begin{table}[]

% \newpage

% The extraction of multiple fields from lengthy, domain-specific incident reports using off‑the‑shelf LLMs (without additional model training) presents significant challenges. 
% To address this, we leverage prompt-based learning techniques, where each LLM extracts information from input data guided by a prompt in runtime. 
Each LLM extracts information from input data guided by a prompt, which significantly affects the accuracy and cost of extraction results.
To design accurate and effective prompts, we use the 
  \emph{Chain of Thought (CoT)} and \emph{In-Context Learning (ICL)} prompting techniques, which have  increasingly become popular in LLM downstream applications~\cite{10.1145/3560815}.
% \todo[size=\tiny]{1-2 citations}.
CoT prompts guide LLMs to reason step by step, yielding more accurate answers~\cite{DBLP:conf/nips/KojimaGRMI22, DBLP:conf/nips/Wei0SBIXCLZ22}. In contast, ICL provides a few demonstration examples directly within the input prompt, enabling LLMs to %effectively 
perform diverse downstream NLP tasks~\cite{OpenAI-ICL, DBLP:conf/nips/BrownMRSKDNSSAA20}.

\textit{Prompt components}: To investigate how different prompt components influence report data extraction, we define five components inspired by prior work~\cite{DBLP:conf/sigsoft/GoelHSGPBZR24}:
Task description (Task), Chain-of-Thought instruction (CoT), Categorization instruction (Category), In-context examples (Examples), and Answering format (Format). The \textit{Task} description instructs the model to assume the role of a system operator and extract information from cloud incident reports. 
(\textit{CoT}) This instruction provides step-by-step guidance to facilitate reasoning about the information. 
% The \textit{Category} instruction defines the classification scheme of different fields, as shown in \Cref{tab:sub-class}. 
For  in-context \textit{Examples}, two samples from the dataset are provided to illustrate the expected answers. 
Finally, the answering \textit{Format} specifies the JSON structure in which the model should return its output.

\textit{Six prompt strategies}: 
% In our case, extracting multiple fields from lengthy, domain-specific incident reports using off‑the‑shelf LLMs (without additional model training) presents significant challenges.
%%%; using a few examples can prove to be misleading rather than helping LLM-based extraction\todo[size=\tiny]{1-2 citations needed}.
% To address this, 
% we design a prompt template that guides LLMs through step-by-step reasoning using CoT, with the option to incorporate ICL based on a few examples. 
By combining the above introduced prompt components in various ways, we construct six prompting strategies, summarized in \Cref{tab:prompt-strategy}. Among them, Task and Format serve as the foundational elements present in all prompts, while the remaining components are selectively incorporated to evaluate their individual and combined effects.

% CoT prompts guide LLMs to reason step by step to give more accurate answers~\cite{DBLP:conf/nips/KojimaGRMI22, DBLP:conf/nips/Wei0SBIXCLZ22}. 
% %These strategies have become increasingly popular in LLM applications.
% \todo[inline,size=\tiny]{We seem to apply only CoT. Is this correct? Then, we should say _We engineer prompts using CoT but not ICL_. If we apply them alternatingly, we should say _We engineer prompts alternatively using ICL and CoT techniques_.}

% Please add the following requimyred packages to your document preamble:
% \usepackage{multirow}

% \begin{table}[]
\begin{table*}[t]
\centering
\caption{Extraction accuracy~(``Acc.'') of different prompt strategies for GPT-3.5 on the AWS dataset. 
For accuracy metrics EM, TK, BS, see~\Cref{tab:extract-fields}.
For each field, boldface \mygreen{mygreen} text indicates best-performers, and boldface \myred{myred} text indicates worst-performers.}
\vspace{-0.3cm}
\resizebox{0.8\linewidth}{!}{
\begin{tabular}{rcrrrrrr}
\toprule
\textbf{Field / Prompt Strategy}     & \textbf{Acc.}                   & \multicolumn{1}{c}{\textbf{Full-ZS}} & \multicolumn{1}{c}{\textbf{Full-FS}} & \multicolumn{1}{c}{\textbf{Basic-ZS}} & \multicolumn{1}{c}{\textbf{Basic-FS}} & \multicolumn{1}{c}{\textbf{CoT-ZS}} & \multicolumn{1}{c}{\textbf{Categ.-ZS}} \\
\midrule
Service Name                      & \cellcolor{gray!10}EM      & 86.00                                                  & \cellcolor{gray!10}\mygreen{100.00}                         & \myred{52.00}                                                   & \cellcolor{gray!10}\mygreen{100.00}                          & 83.33                                                 & \cellcolor{gray!10}60.67                            \\
Location                          & \cellcolor{gray!10}EM      & 48.00                                                  & \cellcolor{gray!10}\mygreen{96.67}                          & 38.00                                                   & \cellcolor{gray!10}83.33                           & 57.33                                                 & \cellcolor{gray!10}\myred{44.67}                            \\
Start Time                        & \cellcolor{gray!10}EM      & 86.00                                                  & \cellcolor{gray!10}\mygreen{91.33}                          & \myred{70.00}                                                   & \cellcolor{gray!10}89.33                           & 83.33                                                 & \cellcolor{gray!10}72.00                            \\
End Time                          & \cellcolor{gray!10}EM      & 83.33                                                  & \cellcolor{gray!10}\mygreen{86.00}                          & \myred{64.00}                                                   & \cellcolor{gray!10}\mygreen{86.00}                           & 78.67                                                 & \cellcolor{gray!10}71.33                            \\
Timezone                          & \cellcolor{gray!10}EM      & \mygreen{98.67}                                                  & \cellcolor{gray!10}\mygreen{98.67}                          & \mygreen{98.67}                                                  & \cellcolor{gray!10}\mygreen{98.67}                          & \mygreen{98.67}                                                 & \cellcolor{gray!10}\mygreen{98.67}                            \\
Service Categ.                    & \cellcolor{gray!10}EM      & 77.33                                                  & \cellcolor{gray!10}71.33                          & 64.67                                                    & \cellcolor{gray!10}76.0                            & \mygreen{80.00}                                                 & \cellcolor{gray!10}\myred{61.33}                             \\
User Symptom Categ.               & \cellcolor{gray!10}TK      & 88.50                                                  & \cellcolor{gray!10}\mygreen{88.94}                          & 9.76                                                    & \cellcolor{gray!10}50.19                          & \myred{9.33}                                                  & \cellcolor{gray!10}90.00                            \\
User Symptom                      & \cellcolor{gray!10}BS      & 84.33                                                  & \cellcolor{gray!10}92.90                          & 84.79                                                   & \cellcolor{gray!10}\mygreen{93.79}                           & 83.09                                                 & \cellcolor{gray!10}\myred{81.63}                            \\ \midrule
\textbf{Overall}                           & \cellcolor{gray!10}\textbf{Average} & 71.08                                                  & \cellcolor{gray!10}\mygreen{79.23}                          & \myred{49.74}    & \cellcolor{gray!10}73.60                           & 61.44                                                 & \cellcolor{gray!10}62.44                            \\ \bottomrule
\end{tabular}}
\label{tab:acc-prompt}
\vspace*{-0.35cm}
\end{table*}

\textit{An exemplary prompt strategy}: \Cref{fig:prompt-template} shows our full prompt with few-shot examples (Full-FS) for extracting report information.
%, \blue{blue} text indicates extracted content, while \red{red} text indicates information classified by LLMs. 
%
The full prompt incorporates all five components examined in this study: Task, CoT, Category, Examples, and Format. Whereas, the other prompt templates are comparatively less component-rich. 
% All templates are included in our open-sourced artifact, located in the folder \verb|prompts/strategies/|.

% The system prompt gives the LLM a role, guiding how it should interact.
% In the user prompt, we tell the model to extract the information fields by following several steps (labels 1-4), and return them in \verb|JSON| format (label ``Finally'').
% We also include the classification instructions for user symptoms and root causes, as described in \Cref{sec:data:annot}. 

% To evaluate the impact of few-shot examples on extraction performance, we compare two prompting methods: \emph{Zero-shot-CoT (Zero)} only using CoT, and \emph{Few-shot-CoT~(Few)} combining CoT and ICL with \textit{two} examples. 
% In \Cref{fig:prompt-template}, \blue{blue} text indicates extracted content, while \red{red} text indicates information classified by LLMs. 

% \todo[inline,size=\tiny]{What do the distinctive blue and red text in \Cref{fig:prompt-template} represent? Are they two separate examples? Or an example and predicted/generated parts? Or...}

\subsection{Evaluation Metrics}\label{sec:eval-metrics}\label{sec:evaluation-metrics} \label{sec:exp:setup:metrics}
% To compare the performance of  LLMs in our task, we consider the following metrics: \\
\noindent\emph{Accuracy}: 
LLMs extract information for all the fields listed in \Cref{tab:extract-fields}, with accuracy different per field. 
%% We therefore consider accuracy for each field. 
% As summarized in \Cref{tab:extract-fields} per field, the optimal accuracy occurs when the observed accuracy for an LLM matches the ground truth; 
% when it does not, we further quantify the inaccuracy.
%%% \Cref{tab:extract-fields} presents the accuracy evaluation metrics for each extracted field.
We first evaluate the extraction accuracy of \textit{entity} and \textit{class} fields using \emph{Exact Match (EM)}, which measures whether the extracted output matches the labeled ground truth exactly~\cite{rajpurkar-etal-2016-squad}.
% When inaccuracy is observed, 
For the multi-class field \textit{user symptom category}, we use the evaluation measure for classification tasks, which is 
\noindent\emph{Token-level~F1~(TK)}. It computes the harmonic mean of precision and recall based on overlapping tokens ~\cite{10.5555/1394399, DBLP:conf/icse/ShettyBKRN021}.
For the textual fields \textit{user symptoms} and \textit{root causes}, which are sentence-level and require semantic accuracy, therefore, we use \emph{BERTScore~(BS)}~\cite{DBLP:conf/iclr/ZhangKWWA20,DBLP:conf/icse/AhmedGBZZR23} to measure semantic similarity using pre-trained BERT models. 
%%%, which usually apply in text generation and summarization tasks \cite{DBLP:conf/icse/AhmedGBZZR23}. 
% \todo[inline,size=\tiny]{Check if the statement ``When inaccuracy is observed, [we use these additional, quantitative metrics].'' If incorrect, explain differently when we use TK and BS.}

% \noindent
% \emph{Latency}: It measures the time elapsed between model input and output, to understand LLM system-level performance. 

% % \noindent 
% \emph{Token counts and cost}: To evaluate extraction cost, we calculate the \emph{token counts} and the \emph{token price} for each extraction round, including input and output tokens. 
% \input{table/table-acc-prompt}
%\section{Results and Discussion}\label{sec:result}
\section{Performance Analysis of LLMs}\label{sec:result}
% \input{table/table-acc-prompt}

% In this section, we firstly compare accuracy of 6 prompt strategies, and then experimentally analyze model accuracy, using the 6 LLMs and the experiment setup introduced in~\Cref{sec:exp:setup}. We compare the accuracy of the model across different information fields and evaluation metrics, analyze error cases, discuss the performance and cost of each model and prompt strategy, and summarize key insights and implications for the selection of the model and prompt. 
%%% the following info add to README.md
% For \Cref{sec:result:prompt}, experiments results correspond to runs on November 13, 2025.
% For \Cref{sec:result:acc}, \Cref{sec:result:perf-cost}, and \Cref{sec:result:selection}, experiments results correspond to runs on August 27, 2025.
%In each subsection, we first summarize the key findings and then provide a detailed discussion.

\subsection{Comparison of Prompt Strategies}\label{sec:result:prompt}
%To identify the most effective prompting method, %%% BOLD claim of optimality

We present an experimental comparison of the six LLM-prompting strategies introduced in \Cref{sec:exp:setup:prompt}. Each strategy is evaluated on the AWS dataset~(\Cref{tab:dataset}) using GPT-3.5 as the underlying model.
%, and compare the extraction accuracy in \Cref{tab:acc-prompt}. % presents the extraction accuracy of six different prompting strategies. 
\Cref{tab:acc-prompt} presents the accuracy achieved by  each prompting strategy, %for different evaluation metrics (EM, TK, BS) based on \Cref{tab:extract-fields}, 
with per-strategy averages computed using the arithmetic mean reported in the bottom-row.

\begin{itemize}[leftmargin=0cm]
\finding{acc-prompt}{
As expected, the most component-rich strategy, Full-FS, achieves the highest overall accuracy~(79.23\%), and is the best- (single-best-) performer among the six strategies for 6
% (37.5\%) 
of the 8 fields. 
In-context Examples are the most effective single-prompt component: Adding it to Basic-ZS, results in Basic-FS's accuracy of 73.60\%; accuracy increases by nearly 24\% over Basic-ZS and becomes comparable with the more component-rich Full-ZS. Similarly, adding Chain-of-Thought (CoT-ZS) and categorization instructions (Categ-ZS) to Basic-ZS leads to 11.7\% and 12.7\% better accuracy, respectively.
}
\end{itemize}

% Across all useful fields, as expected, among the, six strategies the most component-rich strategy Full-FS provides the most accurate extraction~(79.23\% average accuracy). 
% Surprisingly, Basic-FS ranks second on average~(73.60\%), indicating that in-context examples~(Example) alone substantially improve extraction accuracy, even without adding Chain-of-Thought (CoT) or categorization instructions (Category). 
% Full-ZS performs moderately well (71.08\%), showing that explicit CoT and categorization instructions help in the absence of examples.
% Comparing the accuracy of CoT-ZS~(61.44\%) and Categ.-ZS~(62.44\%) with the Basic-ZS baseline's~(49.74\%), indicates CoT and Category components improve accuracy. %, by 11.7\% and 12.7\%, respectively. 
% The reason is 
% %This demonstrates that 
% CoT and Category instructions guide the model’s reasoning process, and improves its consistency in structured data-extraction tasks.
% The Basic-ZS prompt performs the worst of all strategies, reaching only 49.74\% on average, with lowest accuracy for the service name, start and end time, confirming that component-poor prompts (only Task + Format) are insufficient for complex extraction tasks.

% Overall, our comparison of the six prompting strategies highlights the effectiveness of Examples, CoT, and Category components. In the following, we focus on \textbf{Full-ZS (Zero-shot} and \textbf{Full-FS (Few-shot)}, to further analyze their accuracy, cost, and latency.

% \newpage

\subsection{Extraction Accuracy} \label{sec:result:acc}
\begin{table*}[t]
\centering
\caption{Exact match accuracy on selected (meta-data) fields [\%]. For each row, \mygreen{mygreen} $=$ best (highest), \myred{myred} $=$ worst (lowest).}
\vspace{-0.3cm}
\label{tab:acc-em}\label{tab:exact_match}\label{em-acc}
% First subtable
\begin{subtable}{\linewidth}
\centering
\resizebox{0.9\linewidth}{!}{
\begin{tabular}{rrrrrrrrrrrrr}
\toprule
\multicolumn{1}{c}{}                               & \multicolumn{2}{c}{\textbf{GPT 3.5}}                                 & \multicolumn{2}{c}{\textbf{GPT 4o}}                                  & \multicolumn{2}{c}{\textbf{Claude 3.5}}                              & \multicolumn{2}{c}{\textbf{Claude 4}}                                & \multicolumn{2}{c}{\textbf{Gemini 2.0}}                              & \multicolumn{2}{c}{\textbf{Gemini 2.5}}                              \\
\multicolumn{1}{c}{\multirow{-2}{*}{\textbf{AWS}}} & \multicolumn{1}{c}{\textbf{Zero}} & \multicolumn{1}{c}{\textbf{Few}} & \multicolumn{1}{c}{\textbf{Zero}} & \multicolumn{1}{c}{\textbf{Few}} & \multicolumn{1}{c}{\textbf{Zero}} & \multicolumn{1}{c}{\textbf{Few}} & \multicolumn{1}{c}{\textbf{Zero}} & \multicolumn{1}{c}{\textbf{Few}} & \multicolumn{1}{c}{\textbf{Zero}} & \multicolumn{1}{c}{\textbf{Few}} & \multicolumn{1}{c}{\textbf{Zero}} & \multicolumn{1}{c}{\textbf{Few}} \\ \midrule
Service Name & 84.67 & \cellcolor{gray!10}\mygreen{100.00} & \mygreen{100.00} & \cellcolor{gray!10}\mygreen{100.00} & \myred{76.67} & \cellcolor{gray!10}\mygreen{100.00} & 88.00 & \cellcolor{gray!10}\mygreen{100.00} & 79.33 & \cellcolor{gray!10}\mygreen{100.00} & 98.67 & \cellcolor{gray!10}\mygreen{100.00} \\
Location & \myred{48.00} & \cellcolor{gray!10}96.67 & 83.33 & \cellcolor{gray!10}\mygreen{98.67} & 48.67 & \cellcolor{gray!10}96.67 & 55.33 & \cellcolor{gray!10}97.33 & 76.67 & \cellcolor{gray!10}96.67 & 78.00 & \cellcolor{gray!10}96.00 \\
Start Time & 88.00 & \cellcolor{gray!10}91.33 & 95.33 & \cellcolor{gray!10}\mygreen{96.00} & \myred{84.67} & \cellcolor{gray!10}94.67 & 95.33 & \cellcolor{gray!10}95.33 & 88.67 & \cellcolor{gray!10}94.67 & 95.33 & \cellcolor{gray!10}95.33 \\
End Time & 83.33 & \cellcolor{gray!10}86.00 & 87.33 & \cellcolor{gray!10}\mygreen{88.00} & \myred{66.00} & \cellcolor{gray!10}86.00 & 85.33 & \cellcolor{gray!10}86.67 & 83.33 & \cellcolor{gray!10}\mygreen{88.67} & 86.67 & \cellcolor{gray!10}\mygreen{88.00} \\
Timezone & 98.67 & \cellcolor{gray!10}98.67 & 98.00 & \cellcolor{gray!10}98.67 & \myred{97.33} & \cellcolor{gray!10}98.00 & 98.67 & \cellcolor{gray!10}98.67 & \myred{97.33} & \cellcolor{gray!10}98.67 & 98.67 & \cellcolor{gray!10}98.67 \\
Service Categ. & 73.33 & \cellcolor{gray!10}73.33 & 90.00 & \cellcolor{gray!10}\myred{68.00} & 85.33 & \cellcolor{gray!10}87.33 & 88.00 & \cellcolor{gray!10}89.33 & 84.00 & \cellcolor{gray!10}86.00 & \mygreen{90.67} & \cellcolor{gray!10}\mygreen{90.67} \\
\midrule
\textbf{Average} & 79.33 & \cellcolor{gray!10}91.00 & 92.33 & \cellcolor{gray!10}91.56 & \myred{76.44} & \cellcolor{gray!10}93.78 & 85.11 & \cellcolor{gray!10}94.56 & 84.89 & \cellcolor{gray!10}94.11 & 91.34 & \cellcolor{gray!10}\mygreen{94.78} \\
\bottomrule
\end{tabular}}
% \caption{AWS.}
\label{subtab:em-aws}
\end{subtable}
\vspace{0cm} % space between subtables

% Second subtable
\begin{subtable}{\linewidth}
\centering
\resizebox{0.9\linewidth}{!}{
\begin{tabular}{rrrrrrrrrrrrr}
\toprule
\multicolumn{1}{c}{}                               & \multicolumn{2}{c}{\textbf{GPT 3.5}}                                 & \multicolumn{2}{c}{\textbf{GPT 4o}}                                  & \multicolumn{2}{c}{\textbf{Claude 3.5}}                              & \multicolumn{2}{c}{\textbf{Claude 4}}                                & \multicolumn{2}{c}{\textbf{Gemini 2.0}}                              & \multicolumn{2}{c}{\textbf{Gemini 2.5}}                              \\
\multicolumn{1}{c}{\multirow{-2}{*}{\textbf{AZURE}}} & \multicolumn{1}{c}{\textbf{Zero}} & \multicolumn{1}{c}{\textbf{Few}} & \multicolumn{1}{c}{\textbf{Zero}} & \multicolumn{1}{c}{\textbf{Few}} & \multicolumn{1}{c}{\textbf{Zero}} & \multicolumn{1}{c}{\textbf{Few}} & \multicolumn{1}{c}{\textbf{Zero}} & \multicolumn{1}{c}{\textbf{Few}} & \multicolumn{1}{c}{\textbf{Zero}} & \multicolumn{1}{c}{\textbf{Few}} & \multicolumn{1}{c}{\textbf{Zero}} & \multicolumn{1}{c}{\textbf{Few}} \\ \midrule
Service Name & \,56.84 & \cellcolor{gray!10}\,63.16 & 55.79 & \cellcolor{gray!10}61.05 & 66.32 & \cellcolor{gray!10}\mygreen{67.37} & 61.05 & \cellcolor{gray!10}55.79 & 56.84 & \cellcolor{gray!10}65.26 & 57.89 & \cellcolor{gray!10}\myred{52.63} \\
Location & 63.16 & \cellcolor{gray!10}67.37 & 65.26 & \cellcolor{gray!10}64.21 & 64.21 & \cellcolor{gray!10}69.47 & 66.32 & \cellcolor{gray!10}\mygreen{70.53} & \myred{52.63} & \cellcolor{gray!10}67.37 & 60.00 & \cellcolor{gray!10}65.26 \\
Start Time & 97.89 & \cellcolor{gray!10}\mygreen{100.00} & 98.95 & \cellcolor{gray!10}94.74 & \myred{67.37} & \cellcolor{gray!10}96.84 & 98.95 & \cellcolor{gray!10}\mygreen{100.00} & 68.42 & \cellcolor{gray!10}\mygreen{100.00} & 98.95 & \cellcolor{gray!10}98.95 \\
End Time & 92.63 & \cellcolor{gray!10}\mygreen{96.84} & 94.74 & \cellcolor{gray!10}93.68 & 65.26 & \cellcolor{gray!10}93.68 & 93.68 & \cellcolor{gray!10}\mygreen{96.84} & \myred{64.21} & \cellcolor{gray!10}93.68 & 95.79 & \cellcolor{gray!10}\mygreen{96.84} \\
Timezone & 97.89 & \cellcolor{gray!10}97.89 & 97.89 & \cellcolor{gray!10}95.79 & 97.89 & \cellcolor{gray!10}97.89 & 97.89 & \cellcolor{gray!10}\mygreen{98.95} & \myred{86.32} & \cellcolor{gray!10}\mygreen{98.95} & \mygreen{98.95} & \cellcolor{gray!10}\mygreen{98.95} \\
Service Categ. & 64.21 & \cellcolor{gray!10}\myred{58.95} & \mygreen{67.37} & \cellcolor{gray!10}\mygreen{67.37} & 63.16 & \cellcolor{gray!10}63.16 & 64.21 & \cellcolor{gray!10}66.32 & 60.00 & \cellcolor{gray!10}65.26 & 62.11 & \cellcolor{gray!10}62.11 \\ 
Root Cause Categ. & \myred{61.05} & \cellcolor{gray!10}63.16 & 67.37 & \cellcolor{gray!10}64.21 & 65.26 & \cellcolor{gray!10}71.58 & 63.16 & \cellcolor{gray!10}66.32 & \myred{61.05} & \cellcolor{gray!10}\mygreen{73.68} & 67.37 & \cellcolor{gray!10}70.53 \\
\midrule
% \textbf{Average} & 78.77 & \cellcolor{gray!10}80.70 & 80.00 & \cellcolor{gray!10}79.47 & 70.70 & \cellcolor{gray!10}81.40 & 80.35 & \cellcolor{gray!10}81.40 & \myred{64.74} & \cellcolor{gray!10}\mygreen{81.75} & 78.95 & \cellcolor{gray!10}79.12 \\
\textbf{Average} & 76.24 & \cellcolor{gray!10}78.20 & 78.20 & \cellcolor{gray!10}77.29 & 69.92 & \cellcolor{gray!10}80.00 & 77.89 & \cellcolor{gray!10}79.25 & \myred{64.21} & \cellcolor{gray!10}\mygreen{80.60} & 77.29 & \cellcolor{gray!10}77.90 \\
\bottomrule
\end{tabular}}
\label{subtab:em-azure}
\end{subtable}
\vspace{0cm} % space between subtables

% Third subtable
\begin{subtable}{\linewidth}
\centering
\resizebox{0.9\linewidth}{!}{
\begin{tabular}{rrrrrrrrrrrrr}
\toprule
\multicolumn{1}{c}{}                               & \multicolumn{2}{c}{\textbf{GPT 3.5}}                                 & \multicolumn{2}{c}{\textbf{GPT 4o}}                                  & \multicolumn{2}{c}{\textbf{Claude 3.5}}                              & \multicolumn{2}{c}{\textbf{Claude 4}}                                & \multicolumn{2}{c}{\textbf{Gemini 2.0}}                              & \multicolumn{2}{c}{\textbf{Gemini 2.5}}                              \\
\multicolumn{1}{c}{\multirow{-2}{*}{\textbf{GCP}}} & \multicolumn{1}{c}{\textbf{Zero}} & \multicolumn{1}{c}{\textbf{Few}} & \multicolumn{1}{c}{\textbf{Zero}} & \multicolumn{1}{c}{\textbf{Few}} & \multicolumn{1}{c}{\textbf{Zero}} & \multicolumn{1}{c}{\textbf{Few}} & \multicolumn{1}{c}{\textbf{Zero}} & \multicolumn{1}{c}{\textbf{Few}} & \multicolumn{1}{c}{\textbf{Zero}} & \multicolumn{1}{c}{\textbf{Few}} & \multicolumn{1}{c}{\textbf{Zero}} & \multicolumn{1}{c}{\textbf{Few}} \\ \midrule
Service Name & 83.72 & \cellcolor{gray!10}\mygreen{89.77} & 85.12 & \cellcolor{gray!10}86.05 & 73.02 & \cellcolor{gray!10}88.37 & 83.72 & \cellcolor{gray!10}87.91 & \myred{59.53} & \cellcolor{gray!10}\mygreen{89.77} & 79.07 & \cellcolor{gray!10}89.30 \\
% location & NaN & \cellcolor{gray!10}NaN & NaN & \cellcolor{gray!10}NaN & NaN & \cellcolor{gray!10}NaN & NaN & \cellcolor{gray!10}NaN & NaN & \cellcolor{gray!10}NaN & NaN & \cellcolor{gray!10}NaN \\
Start Time & \myred{25.58} & \cellcolor{gray!10}37.21 & 47.44 & \cellcolor{gray!10}55.35 & 33.49 & \cellcolor{gray!10}44.65 & \mygreen{64.65} & \cellcolor{gray!10}49.77 & 45.12 & \cellcolor{gray!10}46.98 & 57.67 & \cellcolor{gray!10}47.44 \\
End Time & \myred{32.09} & \cellcolor{gray!10}44.19 & 78.60 & \cellcolor{gray!10}84.19 & 66.98 & \cellcolor{gray!10}80.47 & 85.12 & \cellcolor{gray!10}86.98 & 73.95 & \cellcolor{gray!10}77.67 & 85.12 & \cellcolor{gray!10}\mygreen{88.37} \\
Timezone & 74.42 & \cellcolor{gray!10}77.67 & 77.67 & \cellcolor{gray!10}87.91 & 73.49 & \cellcolor{gray!10}74.42 & \mygreen{89.77} & \cellcolor{gray!10}82.79 & \myred{54.88} & \cellcolor{gray!10}77.67 & 86.98 & \cellcolor{gray!10}88.84 \\
Service Categ. & 61.86 & \cellcolor{gray!10}53.02 & 61.86 & \cellcolor{gray!10}\myred{40.93} & 64.19 & \cellcolor{gray!10}56.74 & 62.33 & \cellcolor{gray!10}61.86 & 62.33 & \cellcolor{gray!10}\mygreen{67.91} & 62.33 & \cellcolor{gray!10}64.19 \\ \midrule
\textbf{Average} & \myred{55.53} & \cellcolor{gray!10}60.37 & 70.14 & \cellcolor{gray!10}70.89 & 62.23 & \cellcolor{gray!10}68.93 & \mygreen{77.12} & \cellcolor{gray!10}73.86 & 59.16 & \cellcolor{gray!10}72.00 & 74.23 & \cellcolor{gray!10}75.63 \\
\bottomrule
\end{tabular}}
\label{subtab:em-gcp}
\end{subtable}
% \vspace{0cm} % space between subtables

\end{table*}

\Cref{tab:exact_match} reports the general accuracy of the exact match between different models and prompt methods.

\begin{itemize}[leftmargin=0cm]
\finding{em-prompt}{Few-shot prompting generally improves accuracy for most fields, with a maximum average improvement of 17.34\%. However, it does not improve performance for category extraction. In some cases, models with examples even perform worse.}
% such as GPT-4o for service category on AWS and GCP
\end{itemize}

% According to \reffinding{test}

Generally, few-shot prompting improves accuracy in most cases. In our case, we observed improvements in 76.67\% (23/30) of model–field combinations for GCP, 75.00\% (27/36) for AWS, and 64.29\% (27/42) for AZURE. However, it is less effective for service category and root category extraction, with some cases even showing notable decreases. For example, GPT-4o shows a 22.00\% drop in service category accuracy on AWS and a 20.93\% drop on GCP. A possible explanation is limited patterns in these classification tasks, which may lead to overfitting when few-shot examples are used with advanced reasoning models such as GPT-4o.
% (\refind{em-prompt})

% \begin{itemize}[leftmargin=0cm]
% \finding{em-dataset}{Extraction accuracy depends on the length and complexity of reports across different datasets, with shorter and less complex reports resulting in higher accuracy.
% % For AWS, the best average accuracy is 94.78\% (Gemini~2.5), whereas for AZURE and GCP, the best accuracies are 80.60\% (Gemini~2.0) and 77.12\% (Claude~4), respectively.
% }  
% \end{itemize}

% The extraction accuracy varies between datasets. For AWS, the best average accuracy is 94.78\% (Gemini~2.5 few-shot), whereas for AZURE and GCP, the best accuracies are 80.60\% (Gemini~2.0 few-shot) and 77.12\% (Claude~4 zero-shot), respectively. This variation can be attributed to differences in the length and complexity of the report. 
% Most models perform poorly in time identification for GCP, with all achieving less than 65\% accuracy for the start time. This is because each GCP report contains multiple timestamps, making temporal reasoning more challenging for LLMs. In contrast, the accuracy of start and end time on AWS is generally high, with all models exceeding 80\% accuracy except for Claude~3.5.
% (\refind{em-dataset})

\begin{itemize}[leftmargin=0cm]
\finding{em-model}{
Lightweight models sometimes perform better than advanced models, especially with few-shot examples.
% Advanced models and few-shot-CoT prompts do not necessarily guarantee better performance.
% Advanced models or few-shot-CoT prompts do not necessarily guarantee better performance.
% For instance, the best average accuracies are 81.75\% for AZURE~(Gemini~2.0 with few-shot-CoT) and 77.12\% for GCP~(Claude Sonnet 4 with zero-shot-CoT).
}  
\end{itemize}
Advanced models do not necessarily ensure better performance. For example, the highest average accuracy for AZURE is 80.60\% with Gemini~2.0 few-shot, compared to 77.90\% for Gemini~2.5 few-shot.
% and 77.12\% for GCP~(Claude Sonnet 4 with zero-shot-CoT).
In terms of average accuracy, for Azure, lightweight models with few-shot learning could outperform advanced models without few-shot learning. In contrast, for GCP, advanced models achieve higher accuracy than general models regardless of few-shot prompting.
This suggests that the optimal choice of model and prompting strategy varies across datasets and is not sufficient to select a single model or a fixed combination. 

\begin{itemize}[leftmargin=0cm]
% \finding{perf-latency}{Few-shot prompting incurs higher cost.}
\finding{perf-latency}{Few-shot-CoT exhibits lower latency, even though it requires 1.5–2× more input tokens, likely due to the shorter output tokens.}
% Although few-shot-CoT prompts require 1.5–2× more input tokens, they may exhibit lower latency because of shorter output tokens.}
% \finding{perf-cost}{In all datasets and models, few-shot prompting incurs better average accuracy and higher cost.}
\end{itemize}

Although few-shot-CoT prompts require 1.5–2× more input tokens, they may exhibit lower latency because the structured output guided by the few-shot examples reduces the number of output tokens, as output tokens are the primary contributor to latency \cite{latency-optimization-openai}. For example, on GCP, Claude~4 few-shot completes in 10.00\,s, reducing latency by 1.08\,s, compared to zero-shot (11.08\,s).
% (\refind{perf-latency})
% Reducing output tokens by 50\% can decrease latency by approximately 50\%. In contrast, reducing input tokens has a much smaller effect, with a 50\% reduction in prompt length typically yielding only a 1–5\% improvement in latency.
\vspace{-0.2cm}

\begin{itemize}[leftmargin=0cm]
\finding{cost}{The most expensive models cost 50–60x more than the least costly models.}
\end{itemize}
\vspace{-0.2cm}

Average costs vary significantly across models, with the most expensive models costing 50–60× more than the least costly ones across datasets. For example, on Azure, Claude~4 few-shot costs 190.54~($10^{-4}$\,\$), which is 61.5× and 44.6× higher than Gemini~2.0 zero-shot and few-shot, and costs only 3.10~($10^{-4}$\,\$) and 4.27~($10^{-4}$\,\$), respectively. However, the Gemini~2.0 few-shot shows higher accuracy and much lower latencies. 
This highlights that lightweight models, such as Gemini~2.0, can achieve competitive or even superior accuracy while incurring lower cost and latency, emphasizing the importance of considering performance–cost tradeoffs when selecting LLMs for practical applications.
% (\refind{cost})

\subsection{Implications for Model Selection}\label{sec:result:selection}

% \begin{figure}[t]
%   \centering
%   \resizebox{\linewidth}{!}{
%   \includegraphics[width=0.3\linewidth]{figure/sec_result/figure_bubble_aws.pdf}}
%   \vspace{-0.7cm}
%   \caption{Model selection trade-off between accuracy and cost. Legend: (x, y) $=$ (model, prompt); 0 $=$ zero-shot, 1 $=$ few-shot.}
%   \label{fig:model-selection}
% \vspace{-0.5cm}  
% \end{figure}

% \vspace{-0.03cm}
\begin{itemize}[leftmargin=0cm]
\finding{model-selection}{Lightweight models, including Gemini~2.0 and GPT~3.5, offer a strong balance of accuracy, cost, and latency across datasets. Advanced models, such as Gemini~2.5 and GPT-4o, can achieve higher accuracy but at substantially higher cost and latency, while Claude need further optimization (e.g. fine-tuning) for effective use.}
\end{itemize}
% \vspace{-0.03cm}

\begin{figure*}[t]
  \centering
  \begin{minipage}{\textwidth}
  \begin{subfigure}[b]{0.33\textwidth}
    \includegraphics[width=\linewidth]{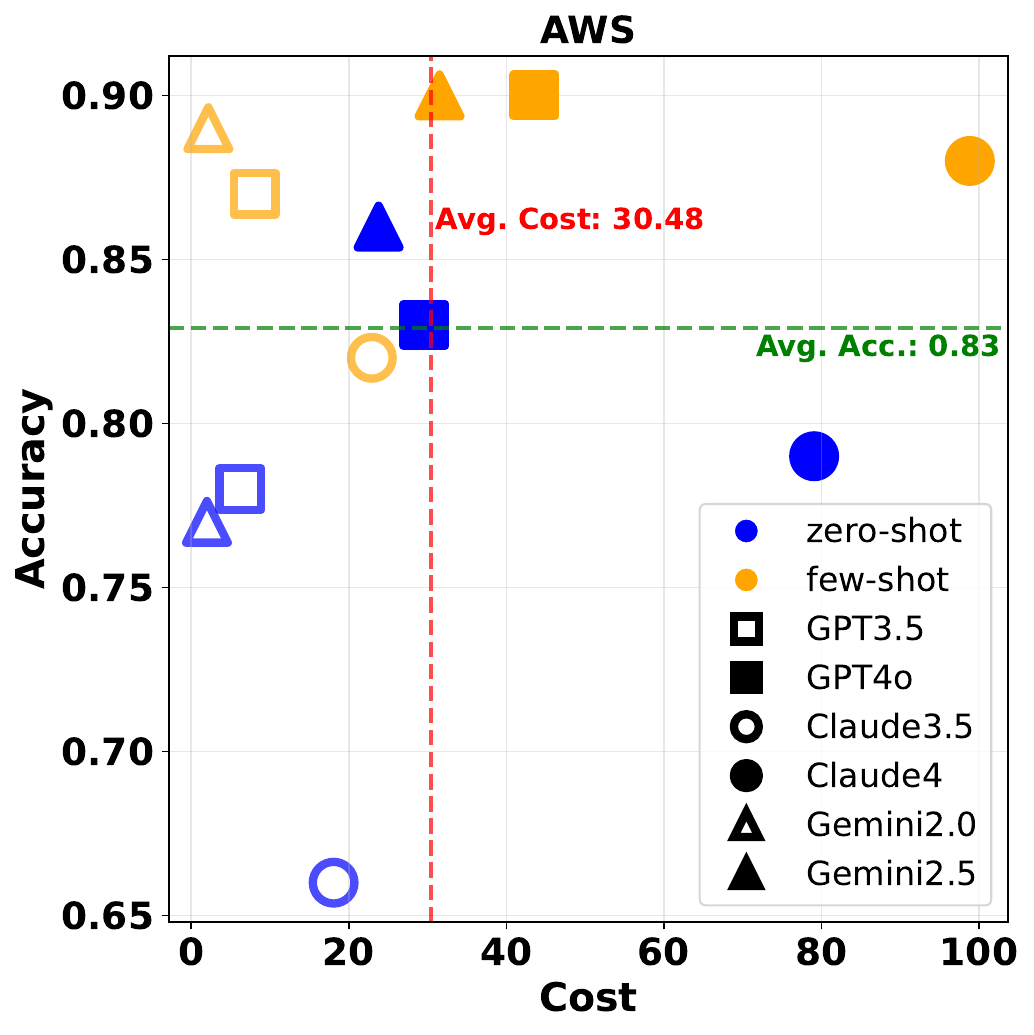}
        \vspace*{-0.6cm}
    \caption{AWS.}
    \label{fig:bubble-aws}
  \end{subfigure}
  \begin{subfigure}[b]{0.33\textwidth}
    \includegraphics[width=\linewidth]{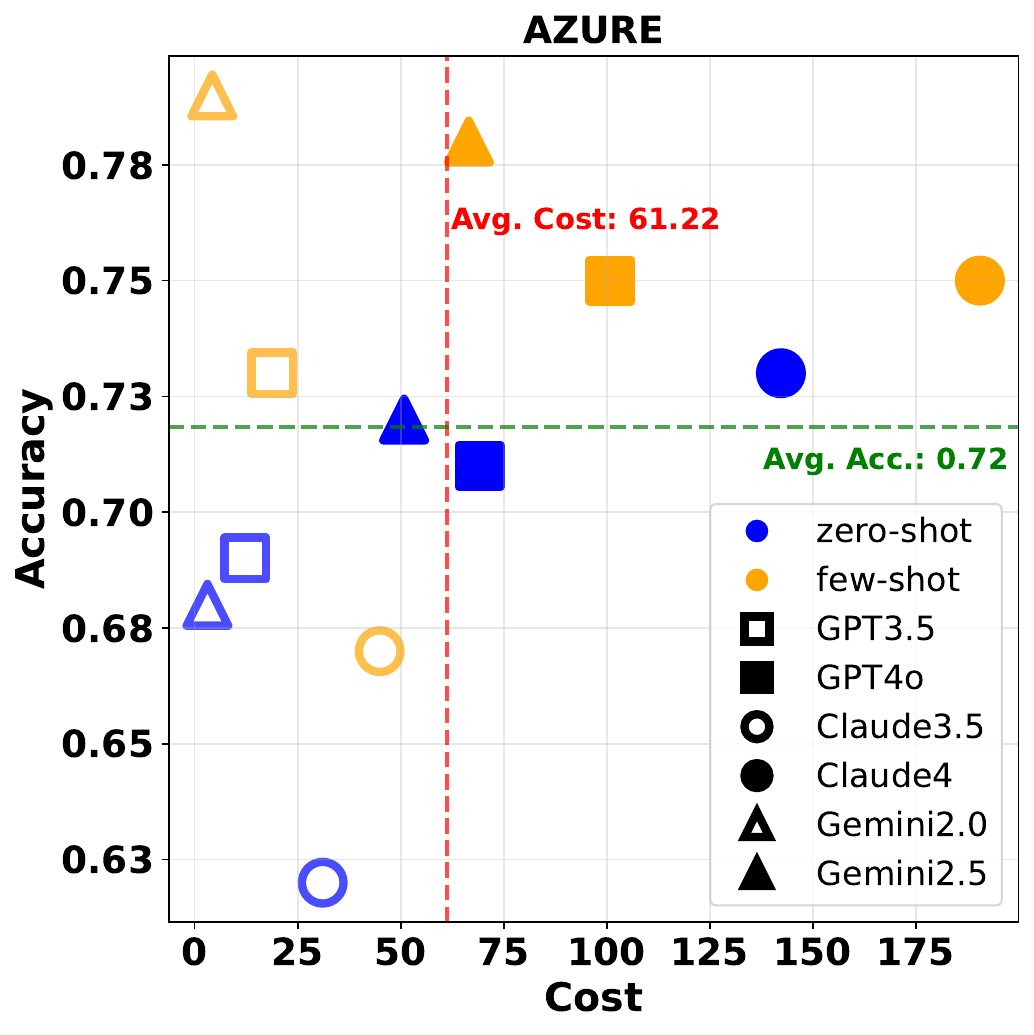}
        \vspace*{-0.6cm}
    \caption{AZURE.}
    \label{fig:bubble-azure}
  \end{subfigure}
\begin{subfigure}[b]{0.33\textwidth}
    \includegraphics[width=\linewidth]{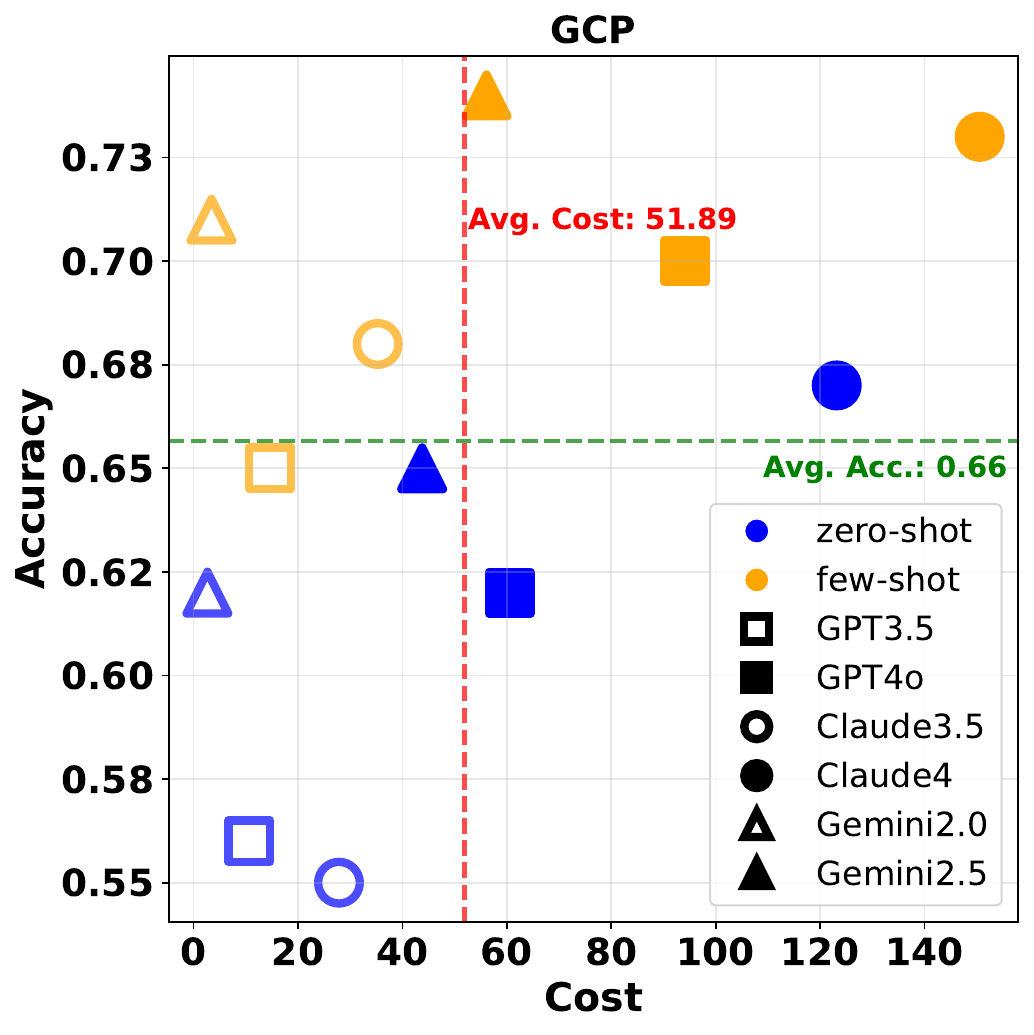}
        \vspace*{-0.6cm}
    \caption{GCP.}
    \label{fig:bubble-gcp}
  \end{subfigure}
  \end{minipage}
  \vspace{-0.4cm}
  \caption{Model selection trade-off between accuracy and cost. }
  % Legend: (x, y) $=$ (model, prompt); 0 $=$ zero-shot, 1 $=$ few-shot.}
  \vspace{-0.3cm}
  \label{fig:model-selection}
\end{figure*}

\Cref{fig:model-selection} illustrates the relationship between accuracy and cost across datasets, models, and prompts. 
% Bubble size represents latency. 
Models positioned toward the upper-left corner indicate higher accuracy at lower cost. 
% Smaller bubbles correspond to lower latency.
The green line represents average accuracy, while the red line represents average cost.
% For AWS, Gemini~2.0 few-shot is the best choice, achieving 0.89 accuracy with lower costs~(2.17) and relatively small latencies~(1.30). If the goal is higher accuracy, Gemini~2.5 or GPT-4o few-shot can be used, though at the expense of slightly higher cost and latency than the average value.
% For Azure, Gemini~2.0 with few-shot outperforms all other options in terms of average accuracy~(0.79), cost~(4.27), and latency~(1.34).
% For GCP, Gemini~2.5 few-shot achieves the highest accuracy~(0.74), but at the expense of approximately 17× higher latency and cost, compared to Gemini~2.0.
Overall, when using LLMs for incident report extraction, we recommend employing few-shot prompts and starting with lightweight models such as Gemini~2.0 and GPT~3.5. Advanced models, such as Gemini~2.5 and GPT-4o, may be considered if higher accuracy is required. Claude series models require further optimization for better application in this task. 

% target: max(accuracy) x min(latency) x min(cost)
% configs: datasets x prompts x models

% \input{sections/5_analysis}

% \newpage
\section{Threats to Validity}\label{sec:threats}

Our work shows how LLMs can be used to extract information from cloud incident reports accurately and effectively. 
However, our approach has certain threats and limitations:

(1) The \emph{accuracy} of the models is constrained by the number and content of the selected few-shot examples. Further research is needed to optimize the design of few-shot prompts for improved accuracy; 
% Moreover, other accuracy evaluation metrics, such as MAE, RMSE, and MSE, could be considered for timestamp data.

(2) The \emph{quality} of the data annotation is limited by the sub-classification of specific fields, such as service category, user symptom category, etc; Also, This "closed-world" classification approach may fail to capture complex and evolving incident types that do not fit neatly into the pre-set buckets, potentially losing critical nuances.

(3) The \emph{depth} of current incident analysis is constrained by the availability of public incident reports. While many numerical results are presented and visually summarized, deeper corelation analysis or causal insights are limited due to the lack of detailed information. For instance, AWS and GCP do not disclose detailed root cause information in their reports, which limits the depth of our analysis. 
% The current incident extraction analysis mainly focuses on descriptive statistics. While many numerical results are presented and visually summarized, deeper corelation analysis or causal insights are limited. The section would benefit from illustrative examples or concrete case studies to better explain the observed patterns.
% \red{add lack of baseline, such as traditional ML/rule-based methods}
\section{Related Work}\label{sec:related-work}

% \textcolor{red}{NA/TODO: table. comparison of different related works}
% Overall, this work complements existing research on incident management and analysis in cloud computing, with a focus on integrating LLM-based extractions. 
% To the best of our knowledge, ours is the first work to systematically explore and evaluate the use of LLMs for structured information extraction and analysis of cloud incident reports, with open-sourced datasets and a toolbox.

\textbf{Incident Management and Analysis:}
% Incident reports and outage studies are essential to understand the reliability of cloud services. 
Cloud incident management and analysis are essential to understand and improve the reliability of cloud computing services, as they provide valuable insights that support root cause detection and mitigation~\cite{7027595, 10.1145/3663529.3663841, DBLP:conf/icse/AhmedGBZZR23}. 
Previous work has been carried out on empirical characterization of outage and incident reports in cloud services~\cite{DBLP:journals/corr/abs-2504-09476, DBLP:conf/cloud/GunawiHSLSAE16, 9659508} , LLM services~\cite{10.1145/3676151.3719372, DBLP:journals/corr/abs-2504-08865}, and frameworks for automation of such analysis~\cite{10.1145/3680256.3721320}. However, these studies either rely on manual or rule-based approaches for information extraction, or focus only on incident metadata fields, without deeper analysis such as user impact and root cause classification. 
% Gunawi et al. \cite{DBLP:conf/cloud/GunawiHSLSAE16} analyzed 597 outages across 32 Internet services, showing that failures persisted despite redundancy due to flaws in detection and recovery. 
% Talluri et al. \cite{DBLP:journals/corr/abs-2504-09476} introduced the Cloud Uptime Archive (CUA), an open dataset combining operator reports, crowdsourced failures, and monitoring data to evaluate fault-tolerance mechanisms. 
% Chu et al. \cite{10.1145/3676151.3719372, 10.1145/3680256.3721320} develop a framework and conduct failure recovery analysis for LLM service incidents. 
% Yan et al. \cite{DBLP:journals/corr/abs-2504-08865} provide the first large-scale study of incidents in GenAI cloud services, revealing distinctive challenges such as higher reliance on human detection and longer mitigation times compared to traditional cloud services. 
The emergence of AI provides solutions for such problems.
There has been research exploring ways to automate the extraction of structured information from incident records using machine learning, such as SoftNER~\cite{DBLP:conf/icse/ShettyBKRN021}, Bayesian networks~\cite{DBLP:conf/www/ChenYLZGXDZDXLK19}, and CUA~\cite{DBLP:journals/corr/abs-2504-09476}. 
However, there is a lack of studies that systematically explore the application of LLMs. 
% but rely mainly on telemetry and internal monitoring rather than public reports. 

% For example, SoftNER \cite{DBLP:conf/icse/ShettyBKRN021} applied unsupervised NER to structure incident records, demonstrating the potential of machine learning for extracting information from unstructured data. Other studies examined broader incident response, including adapting traditional security frameworks to the cloud \cite{9457567} and predicting outages using Bayesian networks and boosting in Microsoft’s systems \cite{DBLP:conf/www/ChenYLZGXDZDXLK19}. 
% These approaches emphasize proactive and reactive management, but rely mainly on telemetry and internal monitoring rather than public reports. 

\textbf{AI for Performance Engineering:}
AI can provide powerful insights into system performance engineering~\cite{10.1145/3676151.3720528}. 
Currently, there are two main application areas that stand out: LLMs for log parsing~\cite{10.1145/3643916.3644408, 10.1145/3674805.3686684}, and LLMs for root cause analysis (RCA)~\cite{DBLP:conf/eurosys/ChenXMKGSCGFWZG24, DBLP:conf/icse/AhmedGBZZR23, 10.1145/3663529.3663841, DBLP:conf/cikm/WangLZZWYF0W24, DBLP:journals/corr/abs-2504-20462}, where LLMs have shown promising performance.
For example, RCACopilot~\cite{DBLP:conf/eurosys/ChenXMKGSCGFWZG24} integrates language models into diagnostic workflows, and TAMO~\cite{DBLP:journals/corr/abs-2504-20462} combines logs, traces, and metrics in a tool-assisted LLM agent to overcome the context and modality limits of RCA. 
However, these works rely mainly on telemetry and internal monitoring data rather than public incident reports. 
Ours is the first study to construct an annotated dataset of incident reports, systematically compare LLMs and prompts for structured information extraction, and perform longitudinal analyses of multiple incident characteristics.

\vspace*{-0.25cm}
\section{Conclusion and Future Work}\label{sec:conclusion}

% Future work:
% \begin{enumerate}
%     \item use RAG to improve extraction accuracy
%     \item Retrieve history report (for finding root cause and predicting failure duration ).
%     \item Hybrid API ensemble approaches.
%     \item Adaptive model selection.
%     \item Mining more information, such as failure-recovery process, fix to the incident.
% \end{enumerate}

% Accurate and efficient data extraction and analysis from cloud incident reports is important for improving the dependability of cloud computing services. 
% Addressing this problem, in this work we present a methodology and reveal the potential to leverage LLM for structure data extraction. %for accurate and efficient data extraction and analysis from cloud incident-reports.
Accurate and efficient data extraction and analysis from cloud incident reports are important for improving the dependability of cloud computing services.
To address this challenge, we propose a methodology that demonstrates how to leverage LLMs for structured data extraction.

In this work, we collect 3,000 incident reports from three cloud operators, and annotate 460 of them for evaluation. 
We propose five prompt components and design six strategies. 
Using both lightweight and advanced LLMs, we then develop data extraction pipeline to extract ten types of information from textual incident reports. 
After that, we evaluate and compare the accuracy, latency, and cost of six LLMs, providing insights into prompt and model selection adapting to different requirements in practical report extraction.
% Finally, we characterize incident duration, service categories, and root cause analyses based on the data extracted by LLMs.
%%% and prompting strategies. 
% Based on our findings, we provide recommendations for selecting models and prompts for practical report extraction. 
Overall, we summarize 6 key findings, and provide open-source artifacts as valuable resources for system researchers, cloud engineers, and service users to better understand and improve cloud incident management. 
% For future work, we aim to apply advanced LLM techniques, such as fine-tuning and Retrieval-Augmented Generation (RAG), to further improve extraction accuracy in specific fields.

Our future work includes: 
(1) Optimized evaluation of prompts. Further research is required to optimize prompt design, and to evaluate different components through controlled experiments.
(2) Proactive incident prediction. The extracted information can be further used to predict incident duration and root causes, which helps better demonstrating downstream utility in incident mitigation.
(3) Advanced LLM techniques. More advanced LLM techniques, such as fine-tuning and~Retrieval Augmented Generation~(RAG), can be applied to further improve extraction accuracy in complex fields thorough historical patterns. 

% (3) Expanded data and analysis. A potential future direction involves integrating these publicly extracted information with internal telemetry or monitoring data to provide a more holistic view of cloud reliability.

%%
%% The acknowledgments section is defined using the "acks" environment
%% (and NOT an unnumbered section). This ensures the proper
%% identification of the section in the article metadata, and the
%% consistent spelling of the heading.
% \newpage
% \section*{Acknoweldgement}
% This work is partially supported by 
% EU MSCA CloudStars (101086248) and 
% Horizon Graph Massivizer (101093202), and 
% by the NL National Growth Fund 6G flagship project Future Network Services.
\begin{acks}
This work is partially supported by 
EU MSCA CloudStars (101086248) and 
Horizon Graph Massivizer (101093202), and 
by the NL National Growth Fund 6G flagship project Future Network Services.
We acknowledge ChatGPT use for grammar and clarity only; this content is original and was written entirely by the authors.
% We acknowledge the support of AI tools, including GPT and DeepSeek, which assisted in refining parts of the manuscript, prior to the final polishing phase that was executed entirely by the authors.
\end{acks}

%%
%% The next two lines define the bibliography style to be used, and
%% the bibliography file.
% \newpage
\bibliographystyle{ACM-Reference-Format}
% \bibliographystyle{abbrvnat}
% \eject
\bibliography{bibliography}

@article{DBLP:journals/corr/abs-2504-09476,
      author={Sacheendra Talluri and Dante Niewenhuis and Xiaoyu Chu and Jakob Kyselica and Mehmet Cetin and Alexander Balgavy and Alexandru Iosup},
  title        = {Cloud Uptime Archive: Open-Access Availability Data of Web, Cloud, and Gaming Services},
  journal      = {CoRR},
  volume       = {abs/2504.09476},
  year         = {2025},
  url          = {https://doi.org/10.48550/arXiv.2504.09476},
  doi          = {10.48550/ARXIV.2504.09476},
  eprinttype    = {arXiv},
  eprint       = {2504.09476},
  bibsource    = {dblp computer science bibliography, https://dblp.org}
}

@inproceedings{10.1145/3676151.3719372,
  author       = {Xiaoyu Chu and
                  Sacheendra Talluri and
                  Qingxian Lu and
                  Alexandru Iosup},
title = {An Empirical Characterization of Outages and Incidents in Public Services for Large Language Models},
year = {2025},
isbn = {9798400710735},
publisher = {Association for Computing Machinery},
address = {New York, NY, USA},
url = {https://doi.org/10.1145/3676151.3719372},
doi = {10.1145/3676151.3719372},
pages = {69–80},
numpages = {12},
location = {Toronto ON, Canada},
series = {ICPE '25}
}

@inproceedings{10.1145/3680256.3721320,
  author       = {S{\'{a}}ndor Battaglini{-}Fischer and
                  Nishanthi Srinivasan and
                  B{\'{a}}lint L{\'{a}}szl{\'{o}} Szarvas and
                  Xiaoyu Chu and
                  Alexandru Iosup},
title = {FAILS: A Framework for Automated Collection and Analysis of LLM Service Incidents},
year = {2025},
isbn = {9798400711305},
publisher = {Association for Computing Machinery},
address = {New York, NY, USA},
url = {https://doi.org/10.1145/3680256.3721320},
doi = {10.1145/3680256.3721320},
booktitle = {Companion of the 16th ACM/SPEC International Conference on Performance Engineering},
pages = {187–194},
numpages = {8},
location = {Toronto ON, Canada},
series = {ICPE '25}
}

@inproceedings{10.1145/3676151.3720528,
author = {John, Lizy K.},
title = {AI for Performance Engineering and Performance Engineering for AI},
year = {2025},
isbn = {9798400710735},
publisher = {Association for Computing Machinery},
address = {New York, NY, USA},
url = {https://doi.org/10.1145/3676151.3720528},
doi = {10.1145/3676151.3720528},
booktitle = {Proceedings of the 16th ACM/SPEC International Conference on Performance Engineering},
pages = {1–2},
numpages = {2},
location = {Toronto ON, Canada},
series = {ICPE '25}
}

@inproceedings{DBLP:conf/sigsoft/GoelHSGPBZR24,
  author       = {Drishti Goel and
                  Fiza Husain and
                  Aditya Singh and
                  Supriyo Ghosh and
                  Anjaly Parayil and
                  Chetan Bansal and
                  Xuchao Zhang and
                  Saravan Rajmohan},
  editor       = {Marcelo d'Amorim},
  title        = {X-Lifecycle Learning for Cloud Incident Management using LLMs},
  booktitle    = {Companion Proceedings of the 32nd {ACM} International Conference on
                  the Foundations of Software Engineering, {FSE} 2024, Porto de Galinhas,
                  Brazil, July 15-19, 2024},
  pages        = {417--428},
  publisher    = {{ACM}},
  year         = {2024},
  url          = {https://doi.org/10.1145/3663529.3663861},
  doi          = {10.1145/3663529.3663861},
  bibsource    = {dblp computer science bibliography, https://dblp.org}
}

@article{10.1145/3560815,
  author       = {Pengfei Liu and
                  Weizhe Yuan and
                  Jinlan Fu and
                  Zhengbao Jiang and
                  Hiroaki Hayashi and
                  Graham Neubig},
title = {Pre-train, Prompt, and Predict: A Systematic Survey of Prompting Methods in Natural Language Processing},
year = {2023},
issue_date = {September 2023},
publisher = {Association for Computing Machinery},
address = {New York, NY, USA},
volume = {55},
number = {9},
issn = {0360-0300},
url = {https://doi.org/10.1145/3560815},
doi = {10.1145/3560815},
journal = {ACM Comput. Surv.},
month = jan,
articleno = {195},
numpages = {35}
}

@article{DBLP:journals/corr/abs-2312-17617,
  author       = {Derong Xu and
                  Wei Chen and
                  Wenjun Peng and
                  Chao Zhang and
                  Tong Xu and
                  Xiangyu Zhao and
                  Xian Wu and
                  Yefeng Zheng and
                  Yang Wang and
                  Enhong Chen},
  title        = {Large Language Models for Generative Information Extraction: {A} Survey},
  journal      = {CoRR},
  volume       = {abs/2312.17617},
  year         = {2023},
  url          = {https://doi.org/10.48550/arXiv.2312.17617},
  doi          = {10.48550/ARXIV.2312.17617},
  eprinttype    = {arXiv},
  eprint       = {2312.17617},
  bibsource    = {dblp computer science bibliography, https://dblp.org}
}

@misc{latency-optimization-openai,
    title={Latency Optimization},
    author={OpenAI API},
    year={2025},
    note={\url{https://platform.openai.com/docs/guides/latency-optimization}, Accessed: 2025-08-18}
}

@inproceedings{10.1145/3674805.3686684,
  author       = {Merve Astekin and
                  Max Hort and
                  Leon Moonen},
title = {A Comparative Study on Large Language Models for Log Parsing},
year = {2024},
isbn = {9798400710476},
publisher = {Association for Computing Machinery},
address = {New York, NY, USA},
url = {https://doi.org/10.1145/3674805.3686684},
doi = {10.1145/3674805.3686684},
booktitle = {Proceedings of the 18th ACM/IEEE International Symposium on Empirical Software Engineering and Measurement},
pages = {234–244},
numpages = {11},
location = {Barcelona, Spain},
series = {ESEM '24}
}

@inproceedings{10.1145/3643916.3644408,
  author       = {Yilun Liu and
                  Shimin Tao and
                  Weibin Meng and
                  Jingyu Wang and
                  Wenbing Ma and
                  Yuhang Chen and
                  Yanqing Zhao and
                  Hao Yang and
                  Yanfei Jiang},
title = {Interpretable Online Log Analysis Using Large Language Models with Prompt Strategies},
year = {2024},
isbn = {9798400705861},
publisher = {Association for Computing Machinery},
address = {New York, NY, USA},
url = {https://doi.org/10.1145/3643916.3644408},
doi = {10.1145/3643916.3644408},
booktitle = {Proceedings of the 32nd IEEE/ACM International Conference on Program Comprehension},
pages = {35–46},
numpages = {12},
location = {Lisbon, Portugal},
series = {ICPC '24}
}

@inproceedings{DBLP:conf/iclr/ZhangKWWA20,
  author       = {Tianyi Zhang and
                  Varsha Kishore and
                  Felix Wu and
                  Kilian Q. Weinberger and
                  Yoav Artzi},
  title        = {BERTScore: Evaluating Text Generation with {BERT}},
  booktitle    = {8th International Conference on Learning Representations, {ICLR} 2020,
                  Addis Ababa, Ethiopia, April 26-30, 2020},
  publisher    = {OpenReview.net},
  year         = {2020},
  url          = {https://openreview.net/forum?id=SkeHuCVFDr},
  bibsource    = {dblp computer science bibliography, https://dblp.org}
}

@book{10.5555/1394399,
author = {Manning, Christopher D. and Raghavan, Prabhakar and Sch\"{u}tze, Hinrich},
title = {Introduction to Information Retrieval},
year = {2008},
isbn = {0521865719},
publisher = {Cambridge University Press},
address = {USA},}

@inproceedings{10.1145/3663529.3663841,
  author       = {Devjeet Roy and
                  Xuchao Zhang and
                  Rashi Bhave and
                  Chetan Bansal and
                  Pedro Henrique B. Las{-}Casas and
                  Rodrigo Fonseca and
                  Saravan Rajmohan},
title = {Exploring LLM-Based Agents for Root Cause Analysis},
year = {2024},
isbn = {9798400706585},
publisher = {Association for Computing Machinery},
address = {New York, NY, USA},
url = {https://doi.org/10.1145/3663529.3663841},
doi = {10.1145/3663529.3663841},
booktitle = {Companion Proceedings of the 32nd ACM International Conference on the Foundations of Software Engineering},
pages = {208–219},
numpages = {12},
location = {Porto de Galinhas, Brazil},
series = {FSE 2024}
}

@inproceedings{rajpurkar-etal-2016-squad,
    title = "{SQ}u{AD}: 100,000+ Questions for Machine Comprehension of Text",
  author       = {Pranav Rajpurkar and
                  Jian Zhang and
                  Konstantin Lopyrev and
                  Percy Liang},
    booktitle = "Proceedings of the 2016 Conference on Empirical Methods in Natural Language Processing",
    month = nov,
    year = "2016",
    address = "Austin, Texas",
    publisher = "Association for Computational Linguistics",
    url = "https://aclanthology.org/D16-1264/",
    doi = "10.18653/v1/D16-1264",
    pages = "2383--2392"
}

@misc{report-aws,
    title={AWS Service Health},
    author={Amazon Web Services},
    year={2025},
    note={\url{https://health.aws.amazon.com/health/status}, Accessed: 2025-09-09}
}

@misc{report-azure,
    title={Azure Status History},
    author={Microsoft Azure},
    year={2025},
    note={\url{https://azure.status.microsoft/en-us/status/history/}, Accessed: 2025-09-09}
}

@misc{report-gcp,
    title={Google Cloud Service Health},
    author={Google Cloud},
    year={2025},
    note={\url{https://status.cloud.google.com/summary}, Accessed: 2025-09-09}
}

@inproceedings{10.1145/3510457.3513030,
author = {Saha, Amrita and Hoi, Steven C. H.},
title = {Mining root cause knowledge from cloud service incident investigations for AIOps},
year = {2022},
isbn = {9781450392266},
publisher = {Association for Computing Machinery},
address = {New York, NY, USA},
url = {https://doi.org/10.1145/3510457.3513030},
doi = {10.1145/3510457.3513030},
booktitle = {Proceedings of the 44th International Conference on Software Engineering: Software Engineering in Practice},
pages = {197–206},
numpages = {10},
location = {Pittsburgh, Pennsylvania},
series = {ICSE-SEIP '22}
}

@inproceedings{DBLP:conf/cloud/GunawiHSLSAE16,
  author       = {Haryadi S. Gunawi and
                  Mingzhe Hao and
                  Riza O. Suminto and
                  Agung Laksono and
                  Anang D. Satria and
                  Jeffry Adityatama and
                  Kurnia J. Eliazar},
  title        = {Why Does the Cloud Stop Computing? Lessons from Hundreds of Service Outages},
  booktitle    = {Proceedings of the Seventh {ACM} Symposium on Cloud Computing, Santa
                  Clara, CA, USA, October 5-7, 2016},
  pages        = {1--16},
  publisher    = {{ACM}},
  year         = {2016},
  url          = {https://doi.org/10.1145/2987550.2987583},
  doi          = {10.1145/2987550.2987583},
  bibsource    = {dblp computer science bibliography, https://dblp.org}
}

@inproceedings{DBLP:conf/nips/Wei0SBIXCLZ22,
  author       = {Jason Wei and
                  Xuezhi Wang and
                  Dale Schuurmans and
                  Maarten Bosma and
                  Brian Ichter and
                  Fei Xia and
                  Ed H. Chi and
                  Quoc V. Le and
                  Denny Zhou},
  title        = {Chain-of-Thought Prompting Elicits Reasoning in Large Language Models},
  booktitle    = {Advances in Neural Information Processing Systems 35: Annual Conference
                  on Neural Information Processing Systems 2022, NeurIPS 2022, New Orleans,
                  LA, USA, November 28 - December 9, 2022},
  year         = {2022},
  url          = {http://papers.nips.cc/paper\_files/paper/2022/hash/9d5609613524ecf4f15af0f7b31abca4-Abstract-Conference.html},
  bibsource    = {dblp computer science bibliography, https://dblp.org}
}

@inproceedings{DBLP:conf/nips/KojimaGRMI22,
  author       = {Takeshi Kojima and
                  Shixiang Shane Gu and
                  Machel Reid and
                  Yutaka Matsuo and
                  Yusuke Iwasawa},
  title        = {Large Language Models are Zero-Shot Reasoners},
  booktitle    = {Advances in Neural Information Processing Systems 35: Annual Conference
                  on Neural Information Processing Systems 2022, NeurIPS 2022, New Orleans,
                  LA, USA, November 28 - December 9, 2022},
  year         = {2022},
  url          = {http://papers.nips.cc/paper\_files/paper/2022/hash/8bb0d291acd4acf06ef112099c16f326-Abstract-Conference.html},
  bibsource    = {dblp computer science bibliography, https://dblp.org}
}

@misc{OpenAI-ICL,
    author = {Alec Radford and
              et al.},
    title = {Language Models are Unsupervised Multitask Learners},
    booktitle = {OpenAI CDN},
    year = {2019},
    url = {https://cdn.openai.com/better-language-models/language_models_are_unsupervised_multitask_learners.pdf}
}

@inproceedings{DBLP:conf/nips/BrownMRSKDNSSAA20,
  author       = {Tom B. Brown and
                  Benjamin Mann and
                  Nick Ryder and
                  Melanie Subbiah and
                  Jared Kaplan and
                  Prafulla Dhariwal and
                  Arvind Neelakantan and
                  Pranav Shyam and
                  Girish Sastry and
                  Amanda Askell and
                  Sandhini Agarwal and
                  Ariel Herbert{-}Voss and
                  Gretchen Krueger and
                  Tom Henighan and
                  Rewon Child and
                  Aditya Ramesh and
                  Daniel M. Ziegler and
                  Jeffrey Wu and
                  Clemens Winter and
                  Christopher Hesse and
                  Mark Chen and
                  Eric Sigler and
                  Mateusz Litwin and
                  Scott Gray and
                  Benjamin Chess and
                  Jack Clark and
                  Christopher Berner and
                  Sam McCandlish and
                  Alec Radford and
                  Ilya Sutskever and
                  Dario Amodei},
  editor       = {Hugo Larochelle and
                  Marc'Aurelio Ranzato and
                  Raia Hadsell and
                  Maria{-}Florina Balcan and
                  Hsuan{-}Tien Lin},
  title        = {Language Models are Few-Shot Learners},
  booktitle    = {Advances in Neural Information Processing Systems 33: Annual Conference
                  on Neural Information Processing Systems 2020, NeurIPS 2020, December
                  6-12, 2020, virtual},
  year         = {2020},
  url          = {https://proceedings.neurips.cc/paper/2020/hash/1457c0d6bfcb4967418bfb8ac142f64a-Abstract.html},
  bibsource    = {dblp computer science bibliography, https://dblp.org}
}

@inproceedings{DBLP:conf/icse/AhmedGBZZR23,
  author       = {Toufique Ahmed and
                  Supriyo Ghosh and
                  Chetan Bansal and
                  Thomas Zimmermann and
                  Xuchao Zhang and
                  Saravan Rajmohan},
  title        = {Recommending Root-Cause and Mitigation Steps for Cloud Incidents using Large Language Models},
  booktitle    = {45th {IEEE/ACM} International Conference on Software Engineering,
                  {ICSE} 2023, Melbourne, Australia, May 14-20, 2023},
  pages        = {1737--1749},
  publisher    = {{IEEE}},
  year         = {2023},
  url          = {https://doi.org/10.1109/ICSE48619.2023.00149},
  doi          = {10.1109/ICSE48619.2023.00149},
  bibsource    = {dblp computer science bibliography, https://dblp.org}
}

@inproceedings{DBLP:conf/eurosys/ChenXMKGSCGFWZG24,
  author       = {Yinfang Chen and
                  Huaibing Xie and
                  Minghua Ma and
                  Yu Kang and
                  Xin Gao and
                  Liu Shi and
                  Yunjie Cao and
                  Xuedong Gao and
                  Hao Fan and
                  Ming Wen and
                  Jun Zeng and
                  Supriyo Ghosh and
                  Xuchao Zhang and
                  Chaoyun Zhang and
                  Qingwei Lin and
                  Saravan Rajmohan and
                  Dongmei Zhang and
                  Tianyin Xu},
  title        = {Automatic Root Cause Analysis via Large Language Models for Cloud
                  Incidents},
  booktitle    = {Proceedings of the Nineteenth European Conference on Computer Systems,
                  EuroSys 2024, Athens, Greece, April 22-25, 2024},
  pages        = {674--688},
  publisher    = {{ACM}},
  year         = {2024},
  url          = {https://doi.org/10.1145/3627703.3629553},
  doi          = {10.1145/3627703.3629553},
  bibsource    = {dblp computer science bibliography, https://dblp.org}
}

@inproceedings{DBLP:conf/www/ChenYLZGXDZDXLK19,
  author       = {Yujun Chen and
                  Xian Yang and
                  Qingwei Lin and
                  Hongyu Zhang and
                  Feng Gao and
                  Zhangwei Xu and
                  Yingnong Dang and
                  Dongmei Zhang and
                  Hang Dong and
                  Yong Xu and
                  Hao Li and
                  Yu Kang},
  editor       = {Ling Liu and
                  Ryen W. White and
                  Amin Mantrach and
                  Fabrizio Silvestri and
                  Julian J. McAuley and
                  Ricardo Baeza{-}Yates and
                  Leila Zia},
  title        = {Outage Prediction and Diagnosis for Cloud Service Systems},
  booktitle    = {The World Wide Web Conference, {WWW} 2019, San Francisco, CA, USA,
                  May 13-17, 2019},
  pages        = {2659--2665},
  publisher    = {{ACM}},
  year         = {2019},
  url          = {https://doi.org/10.1145/3308558.3313501},
  doi          = {10.1145/3308558.3313501},
  bibsource    = {dblp computer science bibliography, https://dblp.org}
}

@inproceedings{DBLP:conf/cikm/WangLZZWYF0W24,
  author       = {Zefan Wang and
                  Zichuan Liu and
                  Yingying Zhang and
                  Aoxiao Zhong and
                  Jihong Wang and
                  Fengbin Yin and
                  Lunting Fan and
                  Lingfei Wu and
                  Qingsong Wen},
  editor       = {Edoardo Serra and
                  Francesca Spezzano},
  title        = {RCAgent: Cloud Root Cause Analysis by Autonomous Agents with Tool-Augmented Large Language Models},
  booktitle    = {Proceedings of the 33rd {ACM} International Conference on Information
                  and Knowledge Management, {CIKM} 2024, Boise, ID, USA, October 21-25,
                  2024},
  pages        = {4966--4974},
  publisher    = {{ACM}},
  year         = {2024},
  url          = {https://doi.org/10.1145/3627673.3680016},
  doi          = {10.1145/3627673.3680016},
  bibsource    = {dblp computer science bibliography, https://dblp.org}
}

@article{DBLP:journals/corr/abs-2504-20462,
  author       = {Qi Wang and
                  Xiao Zhang and
                  Mingyi Li and
                  Yuan Yuan and
                  Mengbai Xiao and
                  Fuzhen Zhuang and
                  Dongxiao Yu},
  title        = {TAMO:Fine-Grained Root Cause Analysis via Tool-Assisted {LLM} Agent with Multi-Modality Observation Data},
  journal      = {CoRR},
  volume       = {abs/2504.20462},
  year         = {2025},
  url          = {https://doi.org/10.48550/arXiv.2504.20462},
  doi          = {10.48550/ARXIV.2504.20462},
  eprinttype    = {arXiv},
  eprint       = {2504.20462},
  bibsource    = {dblp computer science bibliography, https://dblp.org}
}

@inproceedings{DBLP:conf/icse/ShettyBKRN021,
  author       = {Manish Shetty and
                  Chetan Bansal and
                  Sumit Kumar and
                  Nikitha Rao and
                  Nachiappan Nagappan and
                  Thomas Zimmermann},
  title        = {Neural Knowledge Extraction From Cloud Service Incidents},
  booktitle    = {43rd {IEEE/ACM} International Conference on Software Engineering:
                  Software Engineering in Practice, {ICSE} {(SEIP)} 2021, Madrid, Spain,
                  May 25-28, 2021},
  pages        = {218--227},
  publisher    = {{IEEE}},
  year         = {2021},
  url          = {https://doi.org/10.1109/ICSE-SEIP52600.2021.00031},
  doi          = {10.1109/ICSE-SEIP52600.2021.00031},
  bibsource    = {dblp computer science bibliography, https://dblp.org}
}

@article{DBLP:journals/corr/abs-2504-08865,
  author       = {Haoran Yan and
                  Yinfang Chen and
                  Minghua Ma and
                  Ming Wen and
                  Shan Lu and
                  Shenglin Zhang and
                  Tianyin Xu and
                  Rujia Wang and
                  Chetan Bansal and
                  Saravan Rajmohan and
                  Chaoyun Zhang and
                  Dongmei Zhang},
  title        = {An Empirical Study of Production Incidents in Generative {AI} Cloud Services},
  journal      = {CoRR},
  volume       = {abs/2504.08865},
  year         = {2025},
  url          = {https://doi.org/10.48550/arXiv.2504.08865},
  doi          = {10.48550/ARXIV.2504.08865},
  eprinttype    = {arXiv},
  eprint       = {2504.08865},
  bibsource    = {dblp computer science bibliography, https://dblp.org}
}

@INPROCEEDINGS{7027595,
  author       = {Victor Ion Munteanu and
                  Andrew Edmonds and
                  Thomas Michael Bohnert and
                  Teodor{-}Florin Fortis},
  booktitle={2014 IEEE/ACM 7th International Conference on Utility and Cloud Computing}, 
  title={Cloud Incident Management, Challenges, Research Directions, and Architectural Approach}, 
  year={2014},
  volume={},
  number={},
  pages={786-791},
  doi={10.1109/UCC.2014.128}}

@INPROCEEDINGS{9659508,
  author       = {Sacheendra Talluri and
                  Leon Overweel and
                  Laurens Versluis and
                  Animesh Trivedi and
                  Alexandru Iosup},
  booktitle={2021 IEEE International Conference on Autonomic Computing and Self-Organizing Systems (ACSOS)}, 
  title={Empirical Characterization of User Reports about Cloud Failures}, 
  year={2021},
  volume={},
  number={},
  pages={158-163},
  doi={10.1109/ACSOS52086.2021.00039}}

%%
%% If your work has an appendix, this is the place to put it.
% \appendix
% \input{sections/appendix}
% \input{sections/_plan}

\end{document}